\documentclass{llncs}
\usepackage{amssymb}
\usepackage{graphicx}

\usepackage{url}
\urldef{\mails}\path|meizeh@cs.technion.ac.il|

\usepackage{ntheorem}
\theoremseparator{.}

\newtheorem{reducerule}{Reduce}
\newtheorem{branchrule}[reducerule]{Branch}

\newcommand{\comment}[1]{}

\usepackage{algorithm}
\usepackage{algorithmic}
\newcommand{\alg}[1]{\mbox{\sf #1}}  

\usepackage{amsopn}
\DeclareMathOperator{\leaves}{Lea}
\DeclareMathOperator{\internal}{Int}
\DeclareMathOperator{\parent}{par}
\DeclareMathOperator{\children}{Chi}
\DeclareMathOperator{\paths}{Paths}
\DeclareMathOperator{\N}{N}
\DeclareMathOperator{\siblings}{Sib}

\begin{document}

\mainmatter

\title{The $k$-Leaf Spanning Tree Problem\\ Admits a Klam Value of 39\\ (Extended Abstract)}

\author{Meirav Zehavi}

\institute{Department of Computer Science, Technion -- Israel Institute of Technology, Haifa 32000, Israel\\
\mails}

\maketitle

\begin{abstract}
Given an undirected graph $G$ and a parameter $k$, the {\sc $k$-Leaf Spanning Tree ($k$-LST)} problem asks if $G$ contains a spanning tree with at least $k$ leaves. This problem has been extensively studied over the past three decades. In 2000, Fellows {\em et al.}~[FSTTCS'00] explicitly asked whether it admits a klam value of 50. A steady progress towards an affirmative answer continued until 5 years ago, when an algorithm of klam value 37 was discovered. In this paper, we present an $O^*(3.188^k)$-time parameterized algorithm for {\sc $k$-LST}, which shows that the problem admits a klam value of 39. Our algorithm is based on an interesting application of the well-known bounded search trees technique, where the correctness of rules crucially depends on the {\em history} of previously applied rules in a non-standard manner.
\end{abstract}

\section{Introduction}

We study the well-known {\sc $k$-Leaf Spanning Tree ($k$-LST)} problem. Given an undirected graph $G=(V,E)$ and a parameter $k$, it asks if $G$ contains a spanning tree with at least $k$ leaves. Due to its general nature, {\sc $k$-LST} has applications in a variety of areas, including, for example, the design of ad-hoc sensor networks (see \cite{appNetDes1,appNetDes2,appNetDes3}) and computational biology (see, e.g., \cite{appCompBio}). Furthermore, {\sc $k$-LST} is tightly linked to the classic {\sc $k$-Connected Dominated Set ($k$-CDS)} problem. Given an undirected graph $G=(V,E)$ and a parameter $k$, {\sc $k$-CDS} asks if $G$ contains a connected dominating set of size at most $k$, where a connected dominating set is a subset $U\subseteq V$ such that the subgraph of $G$ induced by $U$ is connected and every vertex in $V\setminus U$ is a neighbor of a vertex in $U$. On the one hand, given a spanning tree $T$ with at least $k$ leaves, the set of internal vertices of $T$ forms a connected dominating set of size at most $|V|-k$. On the other hand, given a connected dominating set $S$ of size at most $|V|-k$, one can construct a spanning tree $T$ with at least $k$ leaves (simply attach the vertices in $V\setminus S$ as leaves to a tree spanning the subgraph of $G$ induced by $S$).

Even in restricted settings, it has long been established that {\sc $k$-LST} is NP-hard (see, e.g., \cite{NPhard}). Thus, over the past three decades, {\sc $k$-LST} has been extensively studied in the fields of Parameterized Complexity, Exact Exponential-Time Computation and Approximation. We focus on {\em parameterized algorithms}, which attempt to solve NP-hard problems by confining the combinatorial explosion to a parameter $k$. More precisely, a problem is {\em fixed-parameter tractable (FPT)} with respect to a parameter $k$ if it can be solved in time $O^*(f(k))$ for some function $f$, where $O^*$ hides factors polynomial in the input size.

Table \ref{tab:knownresults} presents a summary of known FPT algorithms for {\sc $k$-LST}.\footnote{The {\sc $k$-LST} algorithm in \cite{KW09} is incorrect (see~web.stanford.edu/$\sim$rrwill/projects.html).} The {\em klam value} of an algorithm that runs in time $O^*(f(k))$ is the maximal value $k$ such that $f(k)<10^{20}$. In 2000, Fellows {\em et al.}~\cite{FMRS00} explicitly asked whether {\sc $k$-LST} admits a klam value of 50. A steady progress towards an affirmative answer continued until 2010, when an algorithm of klam value 37 was discovered
by Binkele-Raible and Fernau \cite{RF10}.

\begin{table}[center]
\centering
\begin{tabular}{|l|c|c|c|}
	\hline
	Reference 		                              & First Published     & Klam Value & Running Time       \\\hline\hline	

  Fellows {\em et al.}~\cite{FL88}	          & 1988 \cite{FL88}    & 0          & FPT                \\\hline			
		
	Bodlaender {\em et al.}~\cite{B89}	        & 1989 \cite{B89}     & 1          & $O^*((17k^4)!)$    \\\hline
	
	Fellows {\em et al.}~\cite{Book99}	        & 1995 \cite{DF95}    & 5          & $O^*((2k)^{4k})$   \\\hline	
	
	Fellows {\em et al.}~\cite{FMRS00}	        & 2000 \cite{FMRS00}  & 17         & $O^*(14.23^k)$     \\\hline

	Bonsma {\em et al.}~\cite{BBW03}	          & 2003 \cite{BBW03}   & 20         & $O^*(9.49^k)$      \\\hline	
	
	Estivill-Castro {\em et al.}~\cite{ECFLR05} & 2005 \cite{ECFLR05} & 22         & $O^*(8.12^k)$      \\\hline	

	Bonsma {\em et al.}~\cite{BBW08}	          & 2008 \cite{BBW08}   & 24         & $O^*(6.75^k)$      \\\hline	

	Kneis {\em et al.}~\cite{KLR11}	            & 2008 \cite{KLR08}   & 33         & $O^*(4^k)$         \\\hline		
	
	Daligault {\em et al.}~\cite{DGKY10}	      & 2008 \cite{DGKY08}  & 35         & $O^*(3.72^k)$      \\\hline		
	
	Binkele-Raible {\em et al.}~\cite{RF14}     & 2010 \cite{RF10}    & 37         & $O^*(3.46^k)$      \\\hline	
	
	{\bf This paper}								            & {\bf 2015}          & {\bf 39}   & $\bf O^*(3.19^k)$  \\\hline
\end{tabular}\smallskip
\caption{Known FPT algorithms for {\sc $k$-LST}.}
\label{tab:knownresults}
\end{table}

In this paper, we present a deterministic polynomial-space FPT algorithm for {\sc $k$-LST} that runs in time $O(3.188^k+\mathrm{poly}(|V|))=O^*(3.188^k)$, which shows that the problem admits a klam value of 39. Our result, like previous algorithms for this problem, is based on the bounded search trees technique (see Section \ref{sec:prelim}): Essentially, when applying a branching rule, we determine the ``role'' of a vertex in $G$---i.e., we decide whether it should be, in the constructed tree, a leaf or an internal vertex (which, in turn, may determine roles of other vertices). Also, along with the constructed tree (to be completed to a spanning tree), we maintain a list of ``floating leaves''---vertices in $G$ that are not yet attached to the constructed tree, but whose role as leaves has been already determined.

Our result makes the following interesting use of the bounded search trees technique: nodes (of a search tree) depend on the {\em history} of nodes that precede them in a non-standard manner---the correctness of many of our reduction and branching rules crucially relies on formerly executed branching rules, particularly on the fact that certain branches considered by them could not lead to the construction of a solution. More precisely, for certain vertices whose role is to be determined, our decision will rely on the fact that there is no solution in which their parents are leaves. Problematic vertices in whose examination we cannot rely on such a fact will be handled by a ``marking'' approach---we will be able to consider our treatment of them as better than it is, since we previously considered the treatment of the vertices that marked them as worse than it is. 

We would like to remark that the crux of our algorithm is the discovery and demonstration of the usefulness of a tool that we call the ``dependency claim'', which describes that dependency of the nodes (of a search tree) on the nodes that precede them. We consider this claim as the foundation on which we build the rules of our algorithm. More precisely, our rules are carefully designed to preserve the correctness of the dependency claim along the search tree, while, most importantly, allowing us to exploit its implications. 

\bigskip
{\noindent\it Organization.} First, in Section \ref{sec:prelim}, we give necessary information on the bounded search trees technique, along with standard definitions and notation. Then, in Section \ref{sec:algorithm}, we describe our algorithm. Due to lack of space, we cannot present the entire set of 39 rules of our algorithm in the extended abstract. Thus, after describing some necessary ingredients (including our measure and the dependency claim), we only present a brief overview of the entire set of rules, after which we do discuss in detail two rules which capture the spirit of our algorithm. The complete set of rules can be found in Appendix \ref{app:rules} (and also at \alg{arxiv.org/abs/1502.07725}). Finally, in Section \ref{sec:conc}, we conclude the paper.

\section{Preliminaries}\label{sec:prelim}

{\it Bounded Search Trees.} Bounded search trees is a fundamental technique in the design of recursive FPT algorithms (see \cite{Book13}). Roughly speaking, to apply this technique, one defines a list of rules of the form \alg{Rule X. [condition] action}, where \alg{X} is the number of the rule in the list. At each recursive call (i.e., a node in the search tree), the algorithm performs the action of the first rule whose condition is satisfied. If by performing an action, the algorithm recursively calls itself at least twice, the rule is a {\em branching rule}, and otherwise it is a {\em reduction rule}. We only consider polynomial-time actions that increase neither the parameter nor the size of the instance, and decrease/simplify at least one of them. Observe that, at any given time, we only store the path from the current node to the root of the search tree (rather than the entire tree).

The running time of the algorithm can be bounded as follows (see, e.g., \cite{BinkeleThesis}). Suppose that the algorithm executes a branching rule where it recursively calls itself $\ell$ times, such that in the $i^\mathrm{th}$ call, the current value of the parameter decreases by $b_i$. Then, $(b_1,b_2,\ldots,b_{\ell})$ is called the {\em branching vector} of this rule. We say that $\alpha$ is the {\em root} of $(b_1,b_2,\ldots,b_{\ell})$ if it is the (unique) positive real root of $x^{b^*} = x^{b^*-b_1} + x^{b^*-b_2} + \ldots + x^{b^*-b_{\ell}}$, where $b^* = \max\{b_1,b_2,\ldots,b_{\ell}\}$. If $b>0$ is the initial value of the parameter, and the algorithm (a) returns a result when (or before) the parameter is negative, and (b) only executes branching rules whose roots are bounded by a constant $c$, then its running time is bounded by $O^*(c^b)$.

\bigskip
{\noindent\it Standard Definitions and Notation.} Given a graph $G=(V,E)$ and a vertex $v\in V$, let $\N(v)$ denote the set of neighbors of $v$ (in $G$, which will be clear from context). Given subsets $S,U\subseteq V$, let $\paths(S,v,U)$ denote the set of paths that start from a vertex in $S$ and end at the vertex $v$, whose {\em internal} vertices belong to $U$ (only). Given a rooted tree $T=(V_T,E_T)$, let $\leaves(T)$, $\internal(T)$ and $\children_i(T)$ denote the leaf-set, the set of internal vertices and the set of vertices with exactly $i$ children in $T$, respectively. Clearly, $\leaves(T)=\children_0(T)$. Given a vertex $v\in V_T$, let $\parent(v)$ and $\siblings(v)$ denote the parent and set of siblings of $v$ (in $T$, which will be clear from context), respectively.\footnote{Recall that vertices $v$ and $u$ are siblings if they have the same parent.}

\section{The Algorithm}\label{sec:algorithm}

Our algorithm, \alg{Alg}, is based on the bounded search trees technique (see Section \ref{sec:prelim}), in which we integrate the ideas mentioned in the introduction.

\bigskip
{\noindent\it Intermediate Instances.}
Each call to \alg{Alg} is associated with an instance $(G=(V,E), T=(V_T,E_T),L,M,F,k)$. Since $G$ and $k$ are always the graph and parameter given as part of the (original) input, we simplify the notation to $(T,L,M,F)$. This corresponds to:

\begin{itemize}
\item A rooted subtree $T$ of $G$.
\item $L$ (``fixed leaves'') and $M$ (``marked leaves'') are disjoint subsets of $\leaves(T)$.
\item $F$ (``floating leaves'') is a subset of $(\leaves(T)\setminus L)\cup(V\setminus V_T)$.
\end{itemize}

\begin{figure}[!ht]\centering
\includegraphics[scale=0.55]{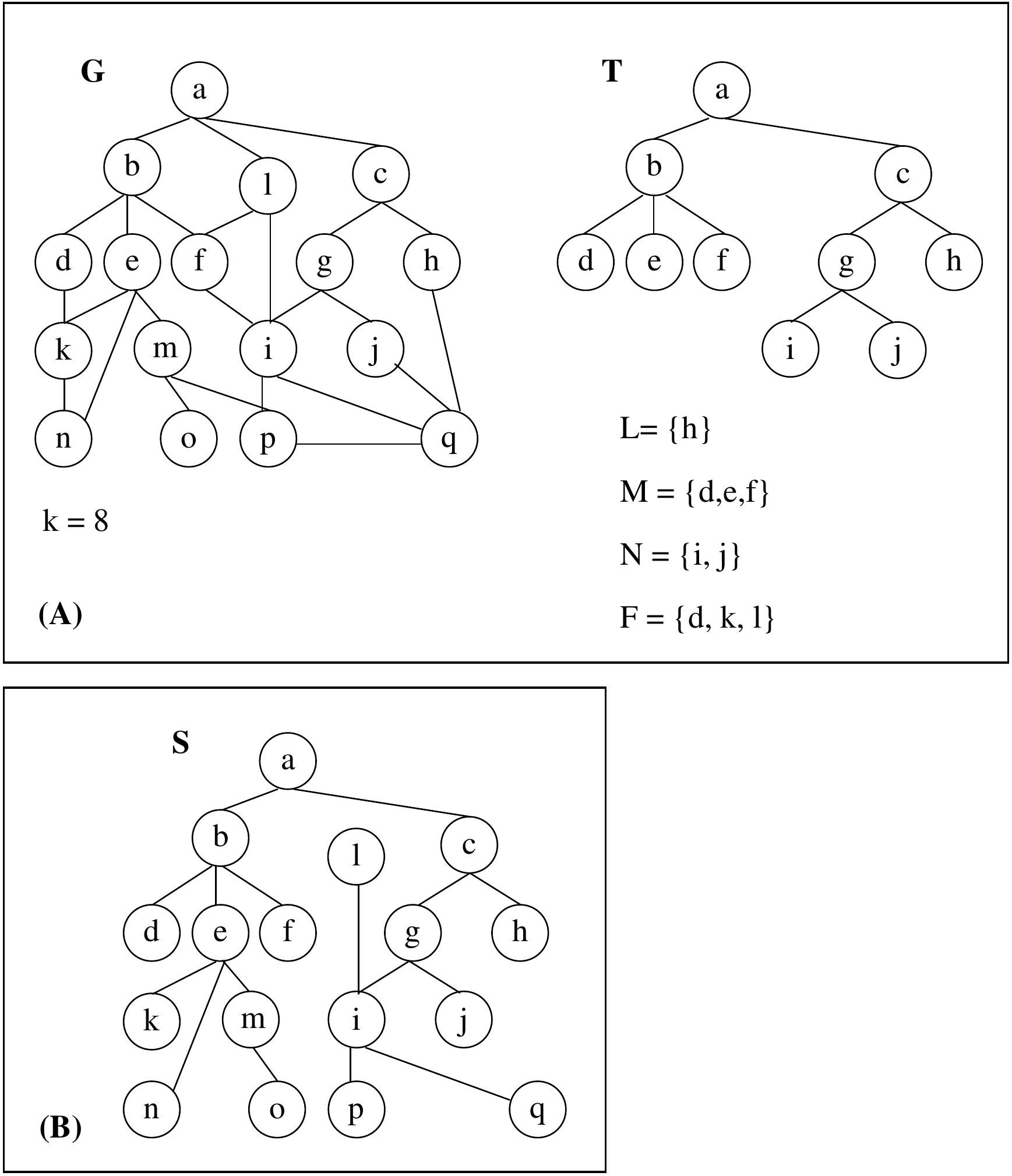}
\caption{(A) An instance associated with a node of the search tree; (B) A spanning tree with 10 leaves that complies with $(T,\{d,h,k,l\})$.}
\label{fig:instance}
\end{figure}

Informally, $T$ is a tree that we try to extend to a solution (that is, a spanning tree of $G$ with at least $k$ leaves) by attaching vertices to its leaves; $L$ contains leaves in $T$ that should be leaves in the solution; $M$ contains leaves in $T$ that other vertices have ``marked'', thus when their roles are decided, the measure (defined below) is decreased by a value large enough for our purpose; $F$ contains vertices in $G$ that are outside $L$, but whose roles as leaves have been already determined. For the sake of clarity, we denote $N=\leaves(T)\setminus(L\cup M)$. That is, $N$ is the set of leaves in $T$ whose roles as leaves (in the solution) has not yet been determined, and which are not marked. An example of an instance $(T,L,M,F)$ is illustrated in Fig.~\ref{fig:instance}(A).

\bigskip
{\noindent\it Goal.} Our goal is to accept the (intermediate) instance {\em iff} $G$ contains a spanning tree $S=(V_S,E_S)$ with at least $k$ leaves that {\em complies with $(T,L\cup F)$}---i.e., (1) $T$ is a subtree of $S$, (2) the vertices in $L\cup F$ are leaves in $S$, and (3) the neighbor set of each internal vertex in $T$ is the same as its neighbor set in $S$. An example of such a spanning tree $S$ is illustrated in Fig.~\ref{fig:instance}(B).

By calling \alg{Alg} with $(T=(\{r\},\emptyset),\emptyset,\{r\},\emptyset)$ for all $r\in V$, and accepting {\em iff} at least one of the calls accepts, we clearly solve {\sc $k$-LST} in time that is bounded by $O^*$ of the running time of \alg{Alg}.

\bigskip
{\noindent\it Measure.} To ensure that the running time of \alg{Alg} is bounded by $O^*(3.188^k)$, we propose the following measure:

\smallskip
{\noindent{\bf Measure:} $\ \displaystyle{2k + \frac{1}{4}|M| -[|L|+|F|+\sum_{i\geq 2}(i-1)|\children_i(T)|]}$.}
\smallskip

Clearly, the measure is initially $2k+\frac{1}{4}$. Moreover, as we will prove in this paper, this measure was carefully chosen to ensure that \alg{Alg} can return a correct decision when the measure drops to (or below) 0, and that the roots of the branching vectors associated with the branching rules are bounded by $3.188^{0.5}$.

We note that at this point, where we have not yet presented our rules, it is already easy to see that the measure makes sense in the following manner: (1) Marking a vertex (i.e., inserting it to $M$) increases the measure, so when the vertex is ``handled'' and thus removed from $M$, its treatment is considered to be better than it actually is, and (2) Determining the role of a vertex as a leaf (i.e., inserting it to $L\cup F$) or an internal vertex with at least two children (i.e., inserting it to $\children_i(T)$ for some $i\geq 2$) decreases the measure by a significant value (at least 1). When determining the role of a vertex as an internal vertex with one child, we can avoid decreasing the measure, since this decision will be made either in a reduction rule or in a branching rule where the role of another vertex, which decreases the measure, is determined.

\bigskip
{\noindent\it The Dependency Claim.} To ensure the correctness of our rules, we will need to preserve the correctness of the {\em dependency claim} (defined below), which describes the dependency of a node in the search tree on the nodes preceding it. This claim supplies information that is relevant only to vertices in $N$, and allows us to handle them as efficiently as we handle marked vertices. More precisely, in some branching rules that determine the roles of vertices in $N$, the dependency claim will justify why certain branches are either omitted or present but also determine the roles of other vertices. Formally, for each $v\in N$, the dependency claim supplies the following information:

\renewcommand{\labelitemi}{$\bullet$}
\renewcommand{\labelitemii}{$-$}
\begin{enumerate}
\item $|\siblings(v)|\leq 1$.
\item Let $(T',L',M',F')$ be the instance associated with the (unique) ancestor node (in the search tree) in which $p\triangleq\parent(v)$ was inserted to $T$ as an internal vertex. Then,
	\begin{enumerate}
	\item There is no solution that complies with $(\widetilde{T},L'\cup F'\cup\{p\})$, where $\widetilde{T}$ is the tree $T'$ from which we remove the descendants of $p$.
	\item If there is $s\in\siblings(v)$, then
		\begin{enumerate}
		\item $s\notin M\cup (\bigcup_{i\geq 2}\children_i(T))$.
		\item $\paths(\leaves(\widetilde{T})\setminus(L'\cup F'\cup\{p\}),s,V\setminus (V_{\widetilde{T}}\cup L'\cup F'))\neq\emptyset$.
		\end{enumerate}
	\end{enumerate}
\end{enumerate}

Roughly speaking, for each vertex $v\in N$, the dependency claim states that we {\em must} have determined (at an ancestor node) that the parent $p$ of $v$ is an internal vertex since otherwise there is no solution ({\bf item 2(a)}). Moreover, it states that if $v$ has a sibling $s$, then $v$ does not have another sibling ({\bf item 1}), the sibling $s$ is neither marked nor has more than one child in $T$ ({\bf item 2(b)i}), and $p$ was not the only vertex from which we could have reached $s$ when we determined the role of $p$ ({\bf item 2(b)ii}).

We note that at this point, where we have not yet presented our rules, it is already possible to see that the dependency claim is potentially useful when handling vertices in $N$. Indeed, suppose that we can use {\bf items 1}, {\bf 2(b)i} and {\bf 2(b)ii} to show in some ``problematic situations'', where the role of at least one vertex in $N$ is determined, that the existence of a solution $S$ where the roles of certain vertices $x,y$ and $z$ are $a,b$ and $c$ implies that there also exists a solution $S'$ which contradicts {\bf item 2(a)}. Then, we can avoid (in advance) setting the roles of $x,y$ and $z$ as $a,b$ and $c$. Having less options to consider implies that \alg{Alg} might examine a smaller search tree, and thus it would be faster.

Observe that initially (i.e., in an instance of the form $(T=(\{r\},\emptyset),\emptyset,\{r\},\emptyset)$), the dependency claim is correct since the root is inserted to $M$.

\bigskip
{\noindent\it Result.} We will show how to devise a set of 39 rules that preserve the correctness of the dependency claim, solve the problem in polynomial-time when the measure drops to (or below) 0, and such that the roots of the branching vectors associated with the branching rules are bounded by $3.188^{0.5}$. We thus obtain that \alg{Alg} runs in time $O^*(3.188^k)$ and uses polynomial-space. Estivill-Castro {\em et al.}~\cite{ECFLR05} proved that in polynomial-time, given an instance $(G=(V,E),k)$ of {\sc $k$-LST}, one can compute another instance $(G'=(V',E'),k)$ of {\sc $k$-LST} such that $|V'|=3.75k$, and $(G,k)$ is a yes-instance {\em iff} $(G',k)$ is a yes-instance. Therefore, by first running the kernelization algorithm in \cite{ECFLR05}, and then calling \alg{Alg} on the instance it computes, we obtain the following result.

\begin{theorem}
The {\sc $k$-LST} problem is solvable in deterministic time $O(3.188^k+\mathrm{poly}(|V|))$, using polynomial-space.
\end{theorem}

\subsection{A Brief Overview of the Rules}\label{sec:overview}

\alg{Alg} starts by examining three (reduction) rules that identify cases where the instance can be solved in polynomial-time. First, Rule 1 rejects the instance when there is a vertex outside the constructed tree $T$ that cannot be attached to $T$ via a path that starts at a vertex in $M\cup N$. Then, Rule 2 shows that when the measure drop to (or below) 0, we have that $k\leq\max\{|\leaves(T)|,|L\cup F|\}$, in which case the instance is necessarily a yes-instance. Now, \alg{Alg} considers the case where all the vertices in $G$ are already contained in $T$---then it concludes (since Rule 2 was not applied) that the instance should be rejected.

Next, \alg{Alg} examines six (reduction) rules that identify cases where the instance, although not necessarily solvable in polynomial-time, is still simple in the sense that we can currently decrease its measure or add a vertex to $T$ without branching. First, Rule 4 turns a floating leaf that is a leaf in $T$ into a fixed leaf. Rule 5 turns a vertex outside $T$ that does not have any neighbor outside $T$ into a floating leaf. Then, Rules 6 and 7 handle {\em certain} situations where there are two vertices such that the neighbor set outside $T$ of one of them is a subset of the neighbor set of the other. In these situations, \alg{Alg} turns one of the two vertices into a floating leaf. Rule 8 handles the case where there is a vertex $v\in M\cup N$ and a vertex $u$ outside $T$ such that $v$ is the only vertex in $M\cup N$ from which we can reach $u$ (without using vertices whose roles have been already determined). In this case, \alg{Alg} determines that $v$ is an internal vertex (while maintaining the correctness of the dependency claim). Rule 9 shows that at this point, if there is a vertex $v\in M\cup N$ that has only one neighbor, $u$, outside the tree, and $u$ also has only one neighbor outside the tree, then $v$ can be turned into a fixed leaf.

Rule 10 considers the case where there is a vertex $v\in M\cup N$ that has exactly two neighbors outside $T$, and these neighbors belong to $F$. It examines two branches to determine the role of $v$ (i.e., to determine whether $v$ should be an internal vertex or a fixed leaf), while relying on the dependency claim to show that if $v$ is an internal vertex, the neighbors of its neighbors in $F$ should be floating leaves.

Now, Rules 11--18 determine the roles of all of the remaining vertices in $M$. First, Rule 11 handles the case where a vertex $v\in M$ has at least three neighbors outside $T$; then, in the branch where it decides that $v$ is an internal vertex, it inserts its children to $M$. Rule 12 handles the case where $v$ has two neighbors outside $T$; then, in the branch where it decides that $v$ is an internal vertex, it shows that its children can be inserted to $N$ without contradicting the dependency claim.\footnote{In this context, recall that we need to avoid marking vertices when it is possible, since each marked vertex increases the measure.} Afterwards, Rule 13 handles the case where $v$ has only one neighbor, $u$, outside $T$, and $u$ has at least three neighbors outside $T$. This rule is similar to Rule 11, where upon deciding that $v$ is internal vertex, $u$ should also be an internal vertex. Rules 14--18 handle the remaining situations, where $v$ has only one neighbor, $u$, outside $T$, and $u$ has two neighbors outside $T$. Here we do not simply perform one rule whose action is similar to the one of Rule 12, since to preserve the correctness of the dependency claim, more delicate arguments are required.

Rules 19--23 handle the situations where there is a vertex $v\in N$ that does not have a sibling in $N$. In the branches of these rules where \alg{Alg} decides that $v$ is a fixed leaf, it relies on the dependency claim to show that it can insert the neighbors outside $T$ of $v$ into $F$. Rule 19 (Rule 20) examines the case where $v$ has at least three (exactly two) neighbors outside $T$. Only when $v$ has exactly two neighbors outside $T$, \alg{Alg} inserts them to $N$ rather than $M$ in the branch where it decides that $v$ is an internal vertex. Then, Rule 21 handles the case where $v$ has only one neighbor, $u$, outside $T$, and $u$ has at least three neighbors outside $T$, while Rules 22 and 23 handle the case where $u$ has two neighbors outside $T$. More precisely, Rule 22 assumes that there is no vertex outside $T$ that can be reached (from a vertex in $N$) only via paths that traverse $u$, while Rule 23 assumes that such a vertex exists. This separation allows \alg{Alg} to perform different actions in Rules 22 and 23 in order to maintain the dependency claim.

Then, Rules 24--28 determine the roles of all the vertices in $N$ that have a sibling in $N$, and this sibling has only one neighbor outside $T$. These rules are similar to rules 19--23, except that now, in order to preserve the correctness of the depndency claim, we also need to handle the sibling, which is simply done by inserting it to $M$. Observe that after Rule 28, we are necessarily handling a situation where there is a vertex $v\in N$ with a sibling $s\in N$, and both $v$ and $s$ have at least two neighbors outside $T$. In most of the following rules, the roles of both $v$ and $s$ are determined together.

Rules 29 and 30 handle the case where there is a vertex $u$ outside $T$ that can be reached only from $v$ and $s$. More precisely, Rule 29 (Rule 30) considers the case where $v$ has at least three (exactly two) neighbors outside $T$. Roughly speaking, these situations require special attention, since upon deciding the roles of $v$ and $s$, it might be problematic to insert their children to $N$ rather than $M$ (for some intuition why this action causes a problem, we refer the reader to item 2(b)ii of the dependency claim). However, rather than considering the situation in these rules as a disadvantage, \alg{Alg} actually exploits it; indeed, for intuition why this is possible, observe that if $v$ is a (fixed) leaf, $s$ must be an internal vertex, since otherwise it is not possible to connect the vertex $u$ to $T$.

Next, Rule 31 handles the case where $v$ and $s$ have a common neighbor outside $T$, and this neighbor is not a floating leaf. The efficiency of this rule relies on an observation (whose correctness is based on the dependency claim) that upon deciding that $v$ is a leaf, \alg{Alg} can also safely decide that $s$ is an internal vertex. After this rule, \alg{Alg} examines two rules, Rules 32 and 33, that handle the (remaining) cases where $v$ has at least three neighbors outside $T$. The separation between Rules 32 and 33 is done according to the number of neighbors in $F$ that $v$ has outside $T$; the case where there are at least two such neighbors is simple, while the case where there is at most one such neighbor requires (in Rule 33) to rely on the depedency claim. 

Finally, in Rules 34--39, \alg{Alg} handles the remaining cases, where both $v$ and $s$ have exactly two neighbors outside $T$. More precisely, Rules 34--36 handle such cases where $v$ and $s$ have a common neighbor outside $T$, and Rules 37--39 handle such cases where the sets of neighbors of $v$ and $s$ outside $T$ are disjoint. The inner distribution of situations corresponding to these cases between Rules 34--36 and between Rules 37--39 is quite delicate, and all of these rules perform actions whose correctness crucially relies on the dependency claim. The following subsection discusses one of these rules in detail.

\subsection{Examples of Central Rules}\label{sec:examples}

In this subsection, we present a reduction rule, as well as a branching rule, that capture the spirit of our algorithm. In particular, Rule 37 demonstrates the power of the dependency claim. We note that each rule is follwed by an illustration.

\setcounter{reducerule}{7}

\begin{reducerule}\label{red:reach}{\normalfont
[There are $v\in M\cup N$ and $u\in V\setminus V_T$ such that $\paths((M\cup N)\setminus\{v\},u,V\setminus (V_T\cup F))=\emptyset$]
Let $X=\N(v)\setminus V_T$.
\begin{enumerate}
\item If $|X|=1$: Return \alg{Alg}$(T'=(V_T\cup X, E_T\cup\{(v,w): w\in X\}),L,M\setminus\{v\},F)$.
\item Return \alg{Alg}$(T'=(V_T\cup X, E_T\cup\{(v,w): w\in X\}),L,(M\setminus\{v\})\cup(\siblings(v)\cap N)\cup X,F)$.
\end{enumerate}
}\end{reducerule}

{\noindent In this rule, there is a vertex $v\in M\cup N$ and a vertex $u$ outside the constructed tree $T$ such that $v$ is the only vertex in $M\cup N$ from which we can reach $u$ (via a path whose internal vertices are neither floating leaves nor belong to $T$). Therefore, if there is a solution $S$ (which, in particular, means that $S$ is a spanning tree that complies with $(T,L\cup F)$), it contains $v$ as an internal vertex. Moreover, the vertices in $X$ are not ancestors of $v$ (since $X\cap V_T=\emptyset$ and $v\in M\cup N$); thus, we can disconnect each of them from its parent in $S$ and attach it to $v$ as a child, obtaining a solution with at least as many leaves as $S$. This implies that we can safely turn $v$ into an internal vertex such that the vertices in $X$ are its children.

In the first case, the measure clearly does not increase. In the second case, the measure both decreases by at least $(|X|-1)$ (since $v$ is inserted to $\children_{|X|}(T)$) and increases by at most $\frac{1}{4}(|X|+1)$ (since $X\cup(\siblings(v)\cap N)$ is inserted to $M$, where by the dependency claim, $|\siblings(v)\cap N|\leq 1$); thus, the measure decreases by at least $\frac{3}{4}|X|-\frac{5}{4}\geq\frac{1}{4}$. Observe that in the second branch, we need to insert $\siblings(v)\cap N$ to $M$, since otherwise we might have a vertex in $N$ whose sibling, $v$, has at least two children in $T$, which contradicts item 2(b)i of the dependency~claim.}

\begin{figure}[!ht]\centering
\frame{\includegraphics[scale=0.6]{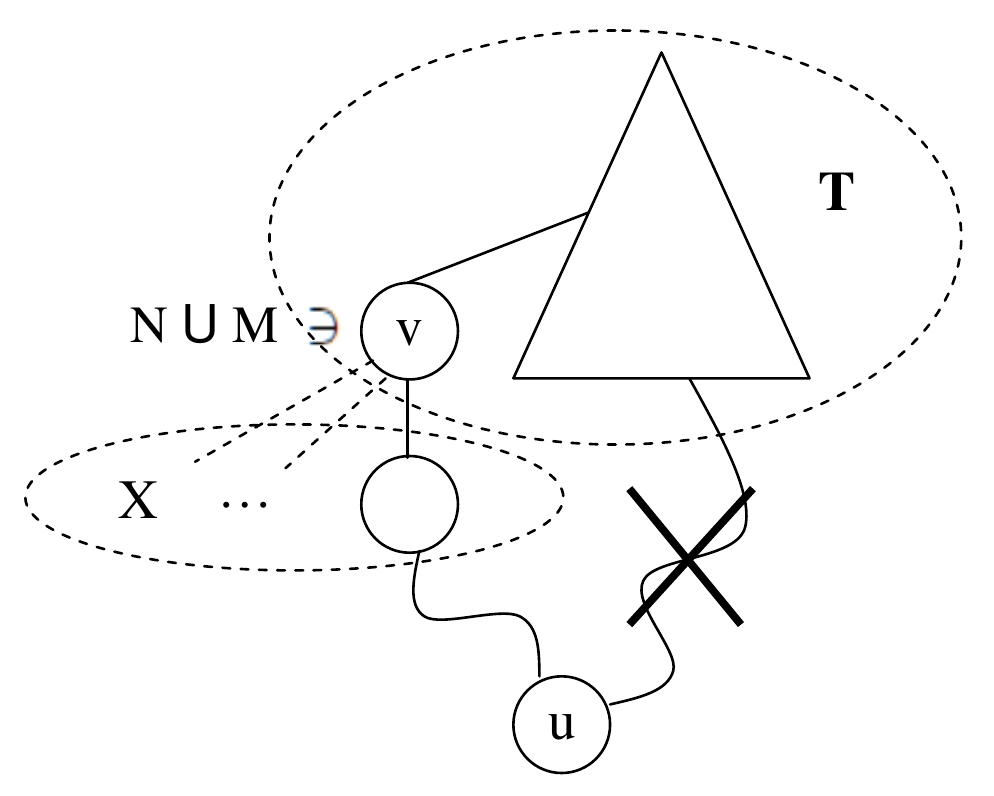}}
\caption{The case handled by Rule \ref{red:reach}.}
\end{figure}

\setcounter{reducerule}{36}

\begin{branchrule}\label{branch:worse}{\normalfont 
[There are $v,s\in N$ such that $s\in\siblings(v)\cap N$, $|X|=|Y|=2$, $X\cap Y=\emptyset$ and $|(X\cup Y)\cap F|\leq 1$, where $X=\N(v)\setminus V_T$ and $Y=\N(s)\setminus V_T$. Moreover, there is no $u\in V\setminus V_T$ such that $\paths(N\setminus\{v,s\},u,V\setminus(V_T\cup F))=\emptyset$]
\begin{enumerate}
\item If \alg{Alg}$(T,L\cup\{v,s\},M,F\cup X\cup Y)$ accepts: Accept.
\item Else if \alg{Alg}$(T'=(V_T\cup X, E_T\cup\{(v,u)\!: u\!\in\! X\}),L\cup\{s\},M,F)$ accepts:~Accept.
\item Else if \alg{Alg}$(T'=(V_T\cup Y, E_T\cup\{(s,u)\!: u\!\in\! Y\}),L\cup\{v\},M,F)$ accepts:~Accept.
\item Return \alg{Alg}$(T'=(V_T\cup X\cup Y, E_T\cup\{(v,u)\!: u\!\in\! X\}\cup\{(s,u)\!: u\!\in\! Y\}),L,M,F)$.
\end{enumerate}
}\end{branchrule}

{\noindent The rule is exhaustive in the sense that we try all four options to determine the roles of $v$ and $s$. Also, recall that once a vertex is determined to be an internal vertex, we can attach each of its neighbors outside $T$ as a child (as explained in Rule \ref{red:reach}). Thus, to prove the correctness of the rule, it suffices to show that in the first branch, inserting the vertices in $X\cup Y$ to $F$ is safe (i.e., if $(T,L\cup\{v,s\},M,F)$ is a yes-instance, then $(T,L\cup\{v,s\},M,F\cup X\cup Y)$ is also a yes-instance). Let $S$ be a solution to $(T,L\cup\{v,s\},M,F)$. Suppose that there is a vertex $u\in X\cap\internal(S)$. Then, we can disconnect (in $S$) the leaf $v$ from its parent and reattach it to $u$, obtaining a spanning tree $S'$ with the same number of leaves as $S$ (since $u\in\internal(S)$). Next, we disconnect the leaf $s$ and reattach it (in $S'$) to another neighbor $w\in V\setminus(\internal(\widetilde{T})\cup L'\cup F')$, where $\widetilde{T}$, $L'$ and $F'$ are defined as in the dependency claim (the existence of $w$ is guaranteed by the dependency claim). We thus obtain a solution $S''$ with at least as many leaves as $S'$, in which the parent of $v$ and $s$ in $T$ is a leaf. By our construction, $S''$ complies with $(\widetilde{T},L'\cup F')$ (since as we progress in a certain branch, we only extend the sets $\internal(T)$ and $L\cup F$). This contradicts the dependency claim. Thus, there is no vertex $u\in X\cap\internal(S)$. Symmetrically, there is no vertex $u\in Y\cap\internal(S)$. Thus, $S$ is also a solution to $(T,L\cup\{v,s\},M,F\cup X\cup Y)$.

Next, we argue that the dependency claim holds in all branches. In the first branch, this is clearly correct. Denote $X=\{x_1,x_2\}$ and $Y=\{y_1,y_2\}$. Now, consider the second branch. There is no solution for $(T,L\cup\{v,s\},M,F)$, since otherwise \alg{Alg} would have accepted in the first branch. Moreover, $x_1,x_2\in\leaves(T')\setminus M$ (where $T'$ is defined in the second branch), and $\paths(N\setminus\{v,s\},x_1,V\setminus (V_{T}\cup F)),\paths(N\setminus\{v,s\},x_2,V\setminus (V_{T}\cup F))\neq\emptyset$ (this follows from the condition of the rule). Therefore, the claim holds in the second branch. Symetrically, the claim holds in the third branch. Similarly, noting that \alg{Alg} did not accept in the second {\em and} third branches, the claim holds in the fourth branch.

Finally, the branching vector is at least as good as $(5,2,2,2)$ since in the first branch, at least five vertices are inserted to $L\cup F$, in each of the second and third branches, one vertex in inserted to $\children_2(T)$ and one vertex is inserted to $L$, and in the fourth branch, two vertices are inserted to $\children_2(T)$. The root of this branching vector is at most $3.188^{0.5}$.}

\begin{figure}[!ht]\centering
\frame{\includegraphics[scale=0.7]{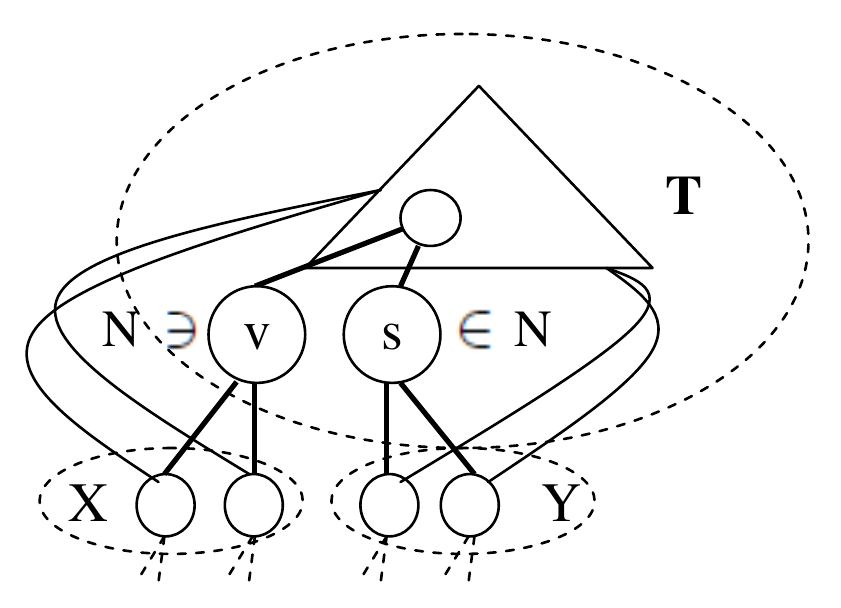}}
\caption{The case handled by Rule \ref{branch:worse}.}
\end{figure}

\section{Conclusion}\label{sec:conc}

In this paper, we developed an $O^*(3.188^k)$-time parameterized algorithm for the {\sc $k$-LST} problem, which implies that it admits a klam value of 39. To this end, we employed the classic bounded search trees technique, in whose application we integrated an interesting claim that captures dependencies between nodes in a compact and useful manner. It is natural to ask whether our approach can be further improved to obtain a faster algorithm for {\sc $k$-LST}, which, in turn, might prove that {\sc $k$-LST} admits a klam value better than 39. Clearly, a more refined set of rules might result in a better running time. However, in order to make significant progress while using our approach, we also suggest to devise a more powerful dependency claim, which should capture more delicate relationships between decisions made along the branches of a search tree.

\bibliographystyle{splncs03}
\bibliography{References}

\newpage

\appendix

\section{The Rules of the Algorithm}\label{app:rules}

In this appendix, we give the rules corresponding to a call \alg{Alg}$(T,L,M,F)$. Each rule is followed by an explanation (including, when relevant, the proof of the preservation of the dependency claim); for a branching rule, we also show that the root is bounded by $3.188^{0.5}$.\footnote{Since our algorithm contains many rules, we find this standard form of presentation clearer than a form where one first lists the rules, and then proves their correctness.} For the sake of clarity, we partitioned the set of rules into smaller subsets corresponding to the description in Section \ref{sec:overview}. Moreover, some rules are followed by illustrations. Although this section contains many rules, once it is understood how to exploit and preserve the dependency claim (in particular, the rules given in Section \ref{sec:examples} demonstrate this), most of the design of \alg{Alg} is quite intuitive.

\subsection{Cases Solvable in Polynomial-Time}

\setcounter{reducerule}{0}

\begin{reducerule}\label{red:stop1}{\normalfont
[There is $v\in (V\setminus V_T)$ s.t.~$\paths(M\cup N,v,V\setminus (V_T\cup F))=\emptyset$]
Reject.
}\end{reducerule}

{\noindent Recall that our goal is to find a solution $S$ (i.e., a spanning tree with at least $k$ leaves) that complies with $(T,L\cup F)$, which means that $T$ is a subtree of $S$, $L\cup F\subseteq\leaves(S)$ and the neighbor set of each internal vertex in $T$ is the same as its neighbor set in $S$. Therefore, if there is a solution $S$ (in the case handled by this rule), it should contain a simple path from a vertex in $M\cup N$ to the vertex $v$ whose internal vertices belong to $V\setminus (V_T\cup F)$. However, $\paths(M\cup N,v,V\setminus (V_T\cup F))=\emptyset$, and therefore such a path does not exist. That is, it is clear that no solution complies with $(T,L\cup F)$, since we cannot connect $v$ to $T$ without using vertices whose roles have been determined (i.e., without using internal vertices, fixed leaves and floating leaves). Therefore, we reject.}

\begin{reducerule}\label{red:stop2}{\normalfont
[$k\leq\max\{|\leaves(T)|,|L\cup F|\}$] 
Accept.
}\end{reducerule}

{\noindent Since the previous rule was not applied, $T$ can be extended to a spanning tree where $L\cup F$ are leaves. Observe that, by extending a tree, we can only obtain a tree with at least as many leaves as the original one. Therefore, $T$ can be extended to a spanning tree with at least $\max\{|\leaves(T)|,|L\cup F|\}$ leaves; thus, if $k\leq\max\{|\leaves(T)|,|L\cup F|\}$, there is a solution, and we accept.}

We now show that if the measure drops to (or below) 0, the algorithm returns a decision, since this rule applies (recall that this statement is necessary to prove that our algorithm runs in the desired time).\footnote{Clearly, \alg{Alg} may return a decision earlier---this can only improve the running time.} Indeed, if $2k + \frac{1}{4}|M| -[|L|+|F|+\sum_{i\geq 2}(i-1)|\children_i(T)|] \leq 0$, we have that $|L\cup F|+\sum_{i\geq 2}(i-1)|\children_i(T)|\geq 2k$. Therefore, either $|L\cup F|\geq k$ or $\sum_{i\geq 2}(i-1)|\children_i(T)|\geq k$. In the former case, the rule clearly applies. Thus, now assume that $\sum_{i\geq 2}(i-1)|\children_i(T)|\geq k$. It is known that for any rooted tree $T'$, $\sum_{i\geq 2}(i-1)|\children_i(T')| = |\leaves(T')| - 2 + \delta$, where $\delta$ is $1$ if the root of $T'$ belongs to $\children_1(T')$ and $0$ otherwise (see, e.g., \cite{kISPbounddeg}). Therefore, we have that $|\leaves(T)|\geq k$, and again, the rule applies.

\begin{reducerule}\label{red:stop3}{\normalfont
[$V=V_T$] 
Reject.
}\end{reducerule}

{\noindent In this rule, $T$ is a spanning tree, and since Rule \ref{red:stop2} was not applied, it contains less than $k$ leaves. Thus, it cannot be extended to a solution, and we reject.}

\subsection{Reduction Rules}

\begin{reducerule}{\normalfont
[There is $v\in \leaves(T)\cap F$] 
Return \alg{Alg}$(T,L\cup\{v\},M\setminus\{v\},F\setminus\{v\})$.
}\end{reducerule}

{\noindent We turn a floating leaf that is a leaf in $T$ into a fixed leaf. The measure does not~increase.}

\begin{reducerule}\label{red:5}{\normalfont
[There is $v\in V\setminus (\internal(T)\cup L\cup F)$ such that $\N(v)\setminus V_T=\emptyset$]
Return \alg{Alg}$(T,L,M\setminus\{v\},F\cup\{v\})$.
}\end{reducerule}

{\noindent We turn a vertex whose role has not yet been determined, and which does not have neighbors outside $T$, into a floating leaf (since it clearly cannot be an internal vertex in a solution). The measure decreases by at least 1.}

\begin{reducerule}\label{red:6}{\normalfont
[There are $v\!\in\! V\!\setminus\!(\internal(T)\!\cup\! L\!\cup\! F)$ and $u\!\in\! M\cup N$ s.t.~$(\N(v)\!\setminus\! V_T)\!\subseteq\!\N(u)$]
Return \alg{Alg}$(T,L,M\setminus\{v\},F\cup\{v\})$.
}\end{reducerule}

{\noindent In this rule, there is a vertex $v$ whose role is undetermined, and whose neighbors outside the tree are also neighbors of some vertex $u\in M\cup N$. Thus, if there is a solution $S$ that contains $v$ as an internal vertex, $v$ is not an ancestor of $u$ (since $u\in M\cup N$) and we can disconnect its children and attach them to $u$, obtaining a solution with at least as many leaves as $S$. Thus, we can safely turn $v$ into a floating leaf. The measure decreases by at least 1.}

\begin{figure}[!h]\centering
\frame{\includegraphics[scale=0.51]{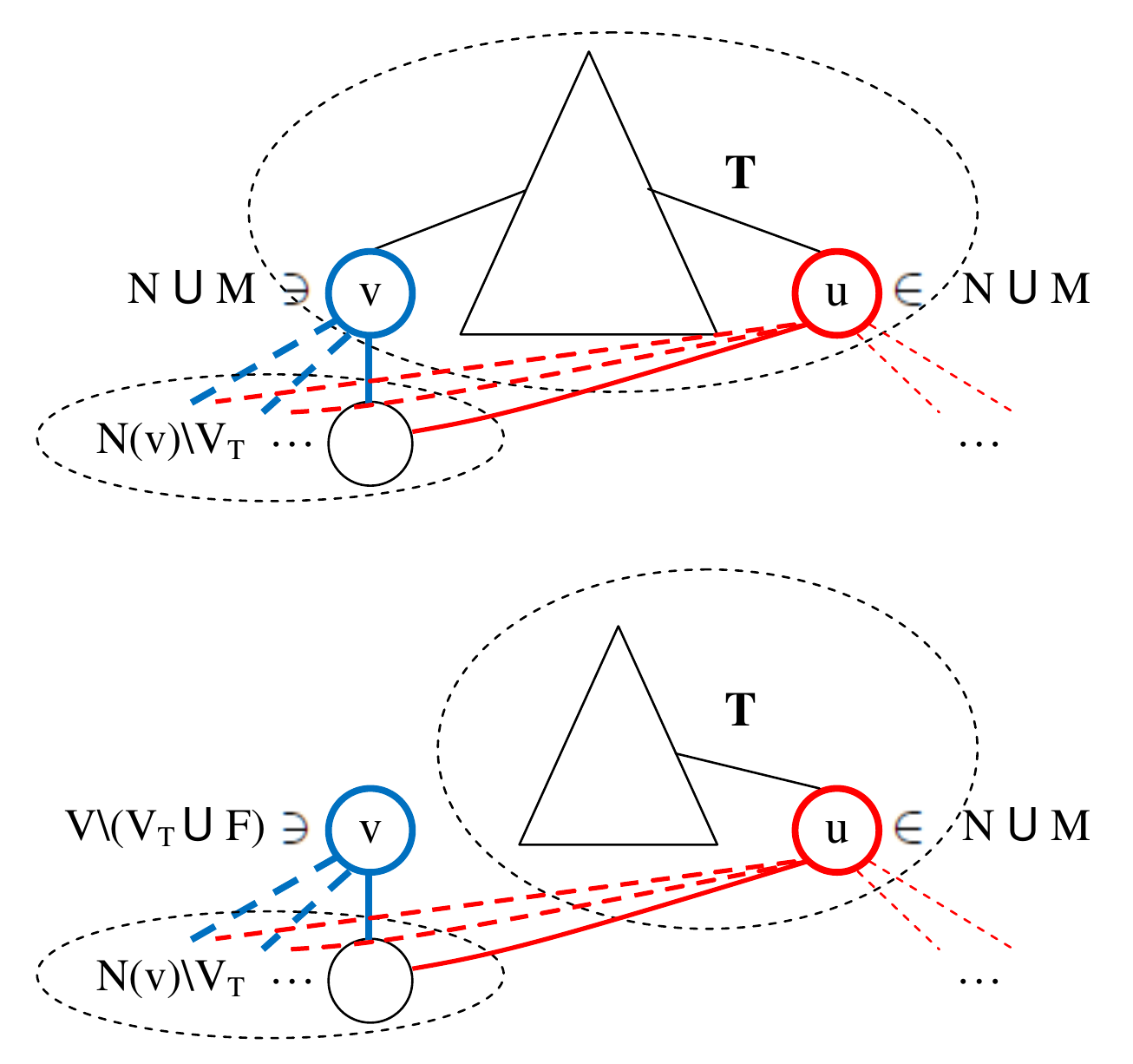}}
\caption{The cases handled by Rule \ref{red:6}.}
\end{figure}

\begin{reducerule}\label{rule:redF}{\normalfont
[There are $v\in V\setminus(\internal(T)\cup L\cup F)$ and $u\in M\cup N$ s.t.~$\N(u)\setminus V_T\subseteq\N(v)$ and $\N(u)\setminus V_T\subseteq F$]
Return \alg{Alg}$(T,L,M\setminus\{u\},F\cup\{u\})$.
}\end{reducerule}

{\noindent In this rule, there is a vertex $v$ whose role is undetermined, along with a vertex $u\in M\cup N$, such that all the neighbors of $u$ outside the tree are floating leaves that are also neighbors of $v$. Thus, if there is a solution $S$ that contains $u$ as an internal vertex, $u$ is not an ancestor of $v$ (since $\N(u)\setminus V_T\subseteq F$) and we can disconnect its children and attach them to $v$, obtaining a solution with at least as many leaves as $S$. Thus, we can safely turn $u$ into a floating leaf. The measure decreases by at least 1.}

\begin{figure}[!h]\centering
\frame{\includegraphics[scale=0.5]{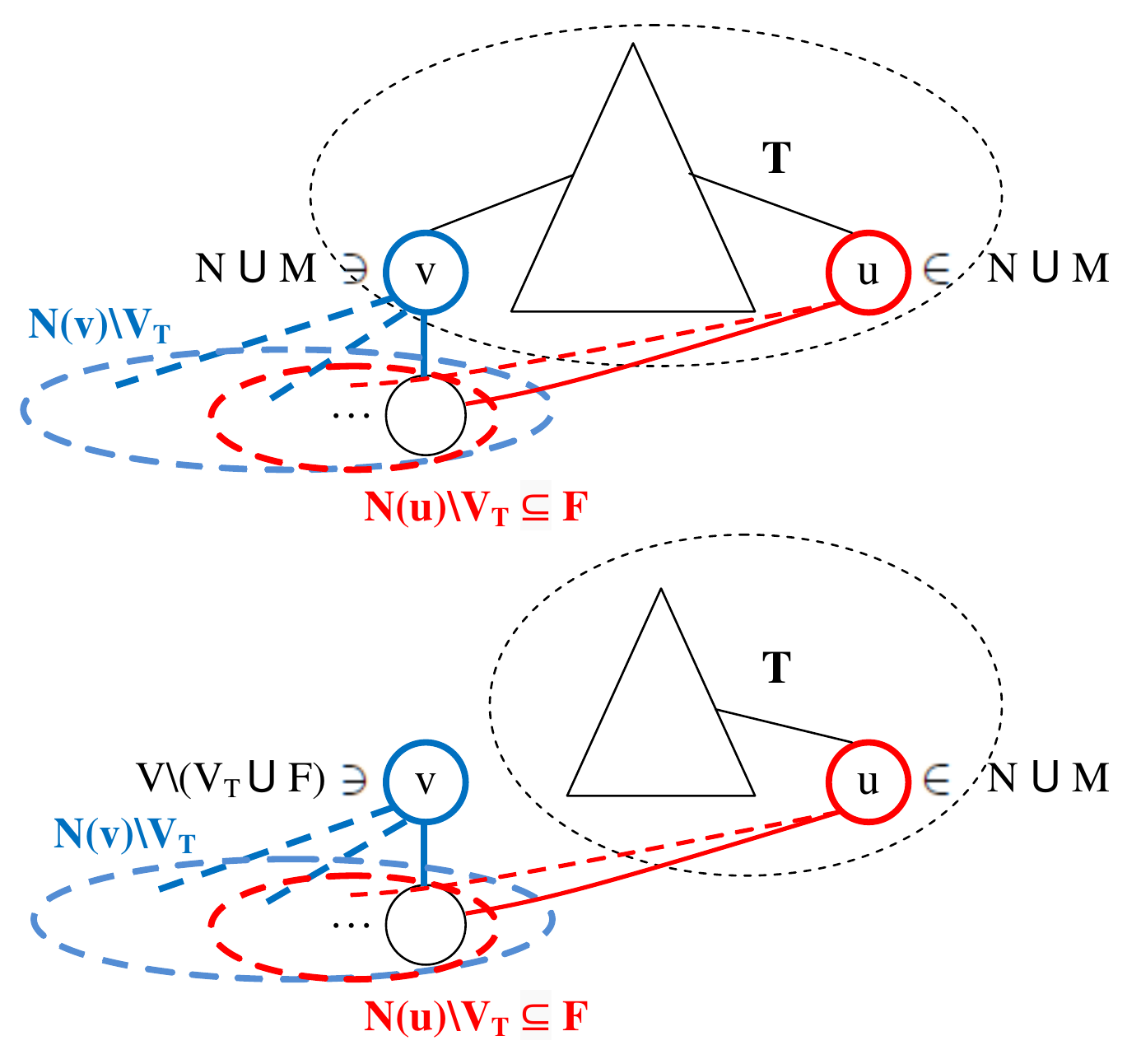}}
\caption{The cases handled by Rule \ref{rule:redF}.}
\end{figure}

\bigskip
{\bf\noindent Rule 8 was given in Section \ref{sec:examples}.}

\setcounter{reducerule}{8}

\begin{reducerule}\label{red:deg2deg2}{\normalfont
[There are $v\in M\cup N$ and $\{u\}=\N(v)\setminus V_T$ such that $|\N(u)\setminus V_T|=1$]
Return \alg{Alg}$(T,L\cup\{v\},M\setminus\{v\},F)$.
}\end{reducerule}

{\noindent In this rule, there is a vertex $v\in M\cup N$ with exactly one neighbor $u$ outside $T$, where $u$ also has only one neighbor outside $T$. Then, if there is a solution $S$ where $v$ is an internal vertex, we can disconnect $u$ from $v$ and attach the subtree of $u$ to another neighbor $w\in V\setminus (\internal(T)\cup L\cup F)$ of a vertex in the subtree (since the previous rule was not applied, a vertex $w$ as required exists), obtaining a solution with at least as many leaves as $S$ (since we turned $v$ into a leaf, and we turned at most two leaves in $S$ into internal vertices---if exactly two, then $u$ was also an internal vertex in $S$ that is now a leaf). Therefore, it is safe to fix $v$ as a leaf. The measure decreases by at least 1.}

\begin{figure}[!ht]\centering
\frame{\includegraphics[scale=0.5]{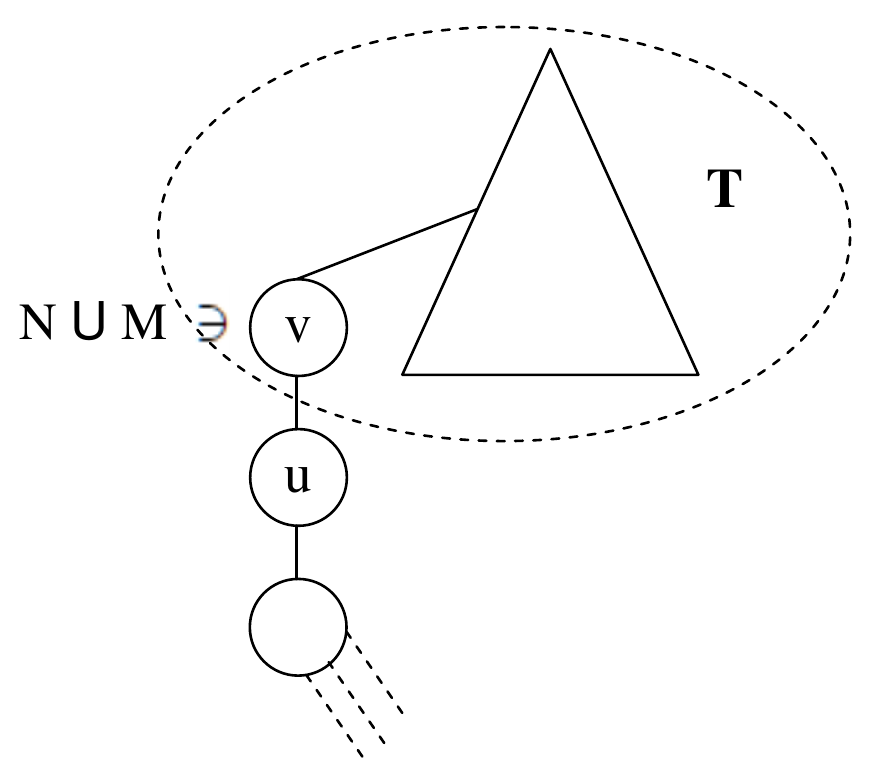}}
\caption{The case handled by Rule \ref{red:deg2deg2}.}
\end{figure}

\subsection{A Leaf Attached to (Two) Floating Leaves}

\begin{branchrule}\label{branch:neisinF}{\normalfont
[There is $v\in M\cup N$ s.t.~$|X|=2$ and $X\subseteq F$, where $X=\N(v)\setminus V_T$]
Let $Z=\N(X)\setminus (\internal(T)\cup L\cup F\cup \{v\})$.\footnote{The standard notation $\N(X)$ refers to the set $(\bigcup_{x\in X}\N(x))\setminus X$.}
\begin{enumerate}
\item If \alg{Alg}$(T,L\cup\{v\},M\setminus\{v\},F)$ accepts: Accept.
\item Return \alg{Alg}$(T'=(V_T\cup X, E_T\cup\{(v,u): u\in X\}),L,(M\setminus\{v\})\cup (\siblings(v)\cap N),F\cup Z)$. 
\end{enumerate}
}\end{branchrule}

{\noindent The rule is exhaustive in the sense that we determine that $v$ is either a leaf or an internal vertex (in which case we can connect the neighbors of $v$ outside $T$ as children of $v$, as explained in Rule \ref{red:reach}). In the second branch, we can assume that there is no solution to $(T,L\cup\{v\},M\setminus\{v\},F)$, and thus, since Rule \ref{red:reach} was not applied, the correctness of inserting $Z$ to $F$ follows in the same manner as the correctness of the insertion of $X\cup Y$ to $F$ in the first branch of Rule \ref{branch:worse} (see Section \ref{sec:examples}). Note that the dependency claim is preserved since $\siblings(v)\cap N$ is inserted to $M$ (in the second branch). Since Rules \ref{rule:redF} and \ref{red:reach} were not applied, $|Z|\geq 2$. Thus, the branching vector, $(1,1+|Z|-\frac{1}{4}|\siblings(v)\cap N|)$, is at least as good as $(1,2\frac{3}{4})$, whose root is smaller than $3.188^{0.5}$. Observe that if $v\in M$, the branching vector is at least as good as $(1\frac{1}{4},3)$.}

\begin{figure}[!ht]\centering
\frame{\includegraphics[scale=0.5]{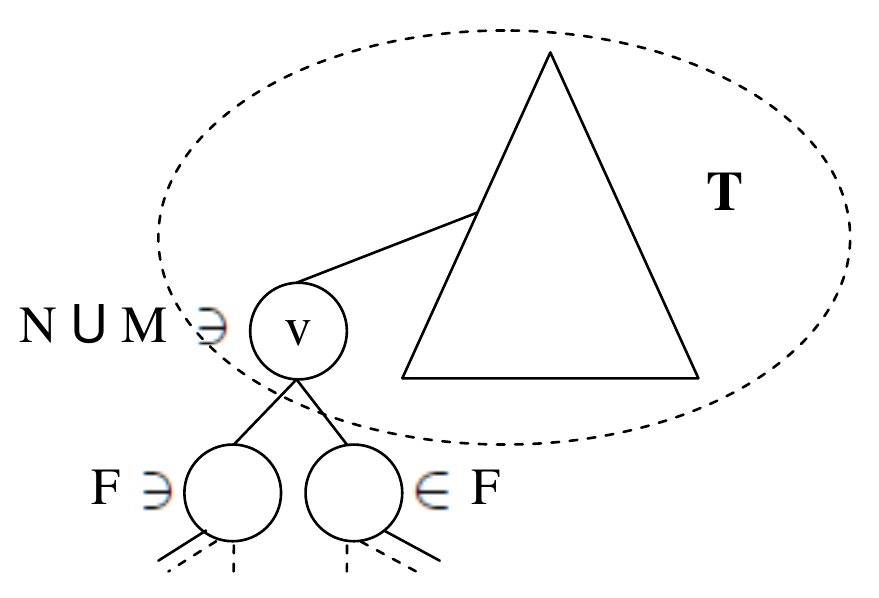}}
\caption{The case handled by Rule \ref{branch:neisinF}.}
\end{figure}

\subsection{Marked Vertices}

\begin{branchrule}\label{rule:M3}{\normalfont
[There is $v\in M$ such that $|X|\geq 3$, where $X=\N(v)\setminus V_T$]
\begin{enumerate}
\item If \alg{Alg}$(T,L\cup\{v\},M\setminus\{v\},F)$ accepts: Accept.
\item Return \alg{Alg}$(T'=(V_T\cup X, E_T\cup\{(v,u): u\in X\}),L,(M\setminus\{v\})\cup (X\setminus F),F)$.
\end{enumerate}
}\end{branchrule}

{\noindent In the first branch, we fix $v$ as a leaf, while in the second branch, we turn $v$ into an internal vertex and attach the vertices in $X$ as its children. Recall that once a vertex is determined to be an internal vertex, we can attach each of its neighbors outside $T$ as a child (as explained in Rule \ref{red:reach}). Since in the second branch, the children of $v$ are inserted to $M$, the dependency claim remains correct.}
 
In the first branch, the measure decreases by $1\frac{1}{4}$ (since $v$ is moved from $M$ to $L$); in the second branch, it decreases by $\frac{1}{4}+(|X|-1)-\frac{1}{4}|X\setminus F|$ (since $v$ is moved from $M$ to $\children_{|X|}(T)$, while the vertices in $X\setminus F$ are inserted to $M$). Thus, the branching vector is $(1\frac{1}{4},\frac{3}{4}(|X|-1))$, which is at least as good as $(1\frac{1}{4},1\frac{1}{2})$, whose root is smaller than $3.188^{0.5}$. Moreover, if $X\subseteq F$, thr branching vector is at least as good as $(1\frac{1}{4},2\frac{1}{4})$.

\begin{figure}[!ht]\centering
\frame{\includegraphics[scale=0.465]{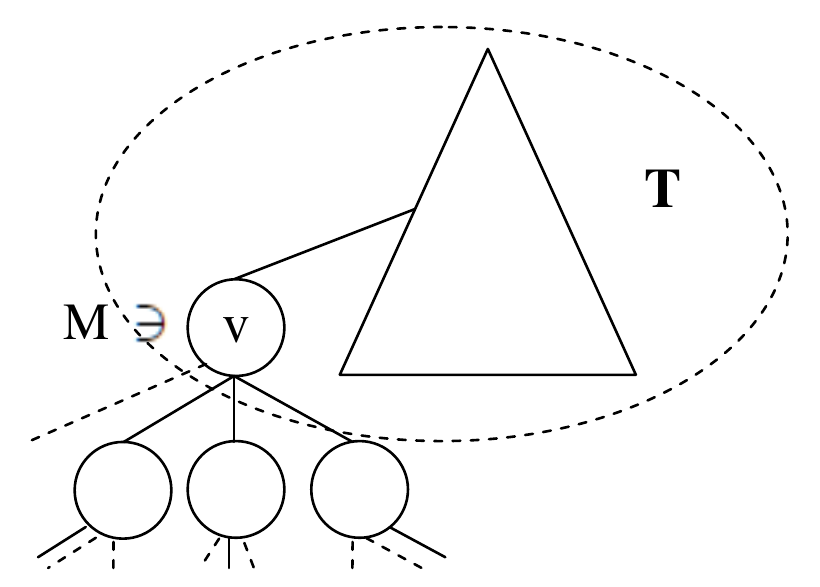}}
\caption{The case handled by Rule \ref{rule:M3}.}
\end{figure}

\begin{branchrule}\label{rule:M2}{\normalfont
[There is $v\in M$ such that $|X|=2$, where $X=\N(v)\setminus V_T$]
\begin{enumerate}
\item If \alg{Alg}$(T,L\cup\{v\},M\setminus\{v\},F)$ accepts: Accept.
\item Return \alg{Alg}$(T'=(V_T\cup X, E_T\cup\{(v,u): u\in X\}),L,M\setminus\{v\},F)$.
\end{enumerate}
}\end{branchrule}

{\noindent For correctness, follow the previous rule, noting that now the vertices in $X$ are inserted to $N$, while the correctness of the dependency claim holds---this is due to the fact that $|X|=2$, Rule \ref{red:reach} was not applied, and the second branch is examined only if the first branch rejected its instance. Since $v$ is moved from $M$ to $L$ (first branch) or $\children_2(T)$ (second branch), the branching vector is $(1\frac{1}{4},1\frac{1}{4})$, whose root is smaller than $3.188^{0.5}$.}

\begin{figure}[!ht]\centering
\frame{\includegraphics[scale=0.5]{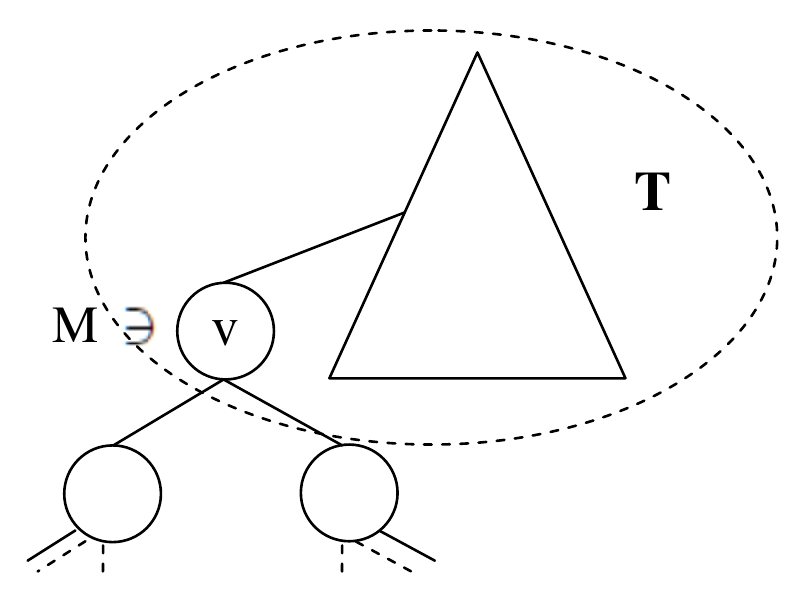}}
\caption{The case handled by Rule \ref{rule:M2}.}
\end{figure}

\begin{branchrule}\label{rule:M33}{\normalfont
[There is $v\in M$ such that $\{u\}=\N(v)\setminus V_T$ and $|X|\geq 3$, where $X=\N(u)\setminus V_T$]
\begin{enumerate}
\item If \alg{Alg}$(T,L\cup\{v\},M\setminus\{v\},F)$ accepts: Accept.
\item Return \alg{Alg}$(T'=(V_T\cup\{u\}\cup X, E_T\cup\{(v,u)\}\cup\{(u,w): w\in X\}),L,(M\setminus\{v\})\cup X,F)$.
\end{enumerate}
}\end{branchrule}

{\noindent This rule is similar to Rule \ref{rule:M3}, only that now, in the second branch where $v$ is turned to an internal vertex, so does $u$. Indeed, if there is a solution $S$ where $v$ is an internal vertex and $u$ is its (only) child that is a leaf, we can disconnect $u$ from $v$ and attach it to another neighbor $w\in V\setminus (\internal(T)\cup L\cup F)$ (since Rule \ref{red:reach} was not applied, a vertex $w$ as required exists), obtaining a solution with at least as many leaves as $S$. Therefore, it suffices to examine (1) $v$ as a leaf, and (2) both $v$ and $u$ as internal vertices. Again, the branching vector is $(1\frac{1}{4},1\frac{1}{2})$, whose root is smaller than $3.188^{0.5}$.}

\begin{figure}[!ht]\centering
\frame{\includegraphics[scale=0.5]{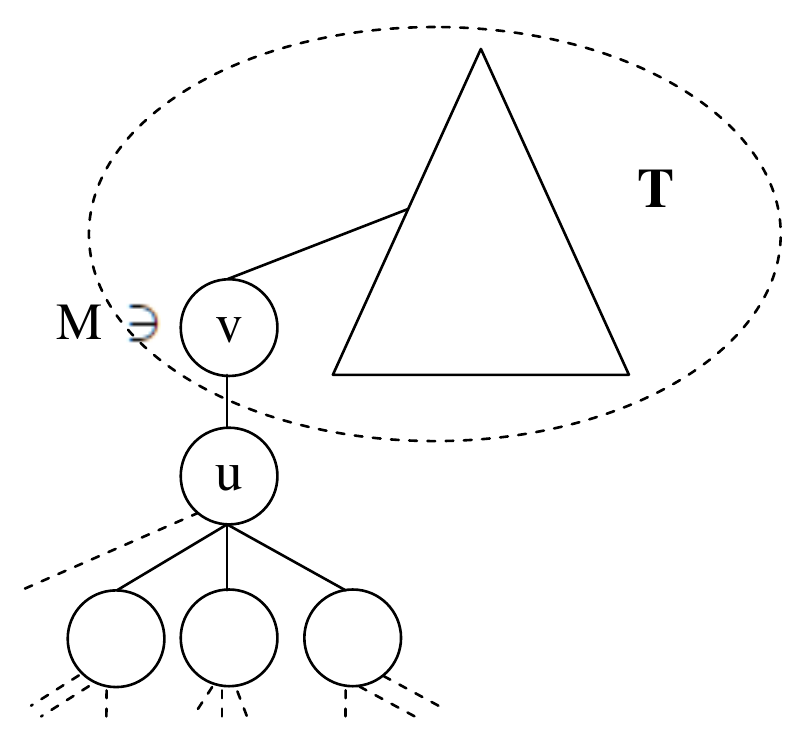}}
\caption{The case handled by Rule \ref{rule:M33}.}
\end{figure}

\begin{branchrule}\label{rule:M2b}{\normalfont
[There is $v\in M$ such that $\{u\}=\N(v)\setminus V_T$, $|X|=2$ and there is no $x\in X$ for which $\paths((M\cup N)\setminus\{v\},x,V\setminus(V_T\cup F\cup\{u\}))=\emptyset$, where $X=\N(u)\setminus V_T$]
\begin{enumerate}
\item If \alg{Alg}$(T,L\cup\{v\},M\setminus\{v\},F)$ accepts: Accept.
\item Return \alg{Alg}$(T'\!=\!(V_T\!\cup\!\{u\}\!\cup\! X, E_T\!\cup\!\{(v,u)\}\!\cup\!\{(u,w): w\in X\}),L,M\setminus\{v\},F)$.
\end{enumerate}
}\end{branchrule}

{\noindent This rule is similar to Rule \ref{rule:M2}, only that now, in the second branch where $v$ is turned to an internal vertex, so does $u$ (an action whose correctness is shown in the previous rule). Noting that there is no $x\in X$ for which $\paths((M\cup N)\setminus\{v\},x,V\setminus(V_T\cup F\cup\{u\}))=\emptyset$, the dependency claim is preserved as in Rule \ref{rule:M2}. Again, the branching vector is $(1\frac{1}{4},1\frac{1}{4})$, whose root is smaller than~$3.188^{0.5}$.}

\begin{figure}[!ht]\centering
\frame{\includegraphics[scale=0.5]{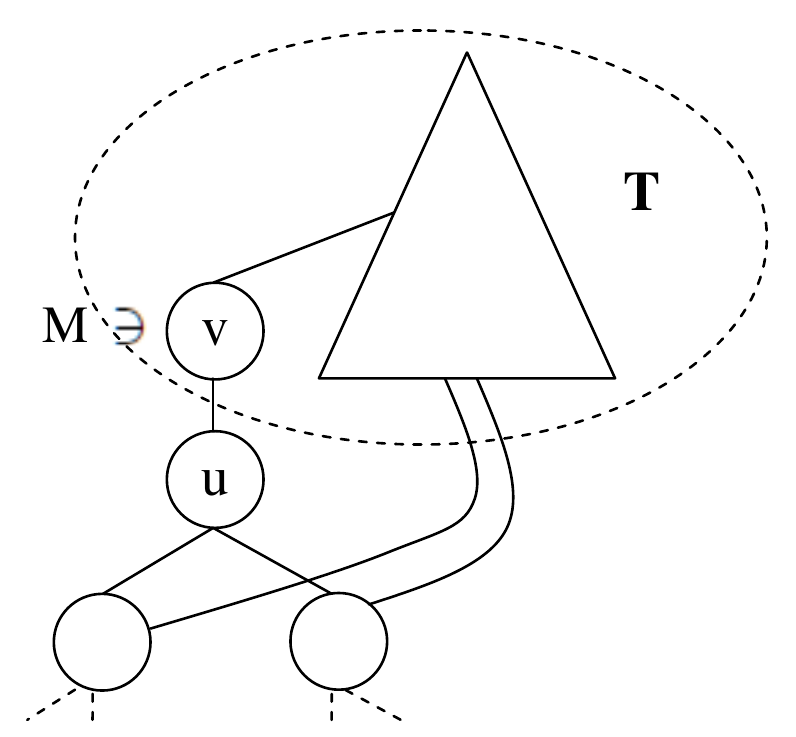}}
\caption{The case handled by Rule \ref{rule:M2b}.}
\end{figure}

\begin{branchrule}\label{branch:Mreduce}{\normalfont
[There is $v'\in M$ such that $\{u'\}=\N(v')\setminus V_T$ and (in the second branch, one of the preceding reduction rules is applicable with $v=x_1$ and then with $v=x_2$), where $\{x_1,x_2\}=\N(u')\setminus V_T$]
\begin{enumerate}
\item If \alg{Alg}$(T,L\cup\{v'\},M\setminus\{v'\},F)$ accepts: Accept.
\item Return \alg{Alg}$(T'=(V_T\cup\{u',x_1,x_2\}, E_T\cup\{(v',u')\}\cup\{(u',x_1),(u',x_2)\}),L,(M\setminus\{v'\})\cup\{x_1,x_2\},F)$.
\end{enumerate}
}\end{branchrule}

{\noindent This rule is similar to the previous one, where now the dependency claim is preserved since $\{x_1,x_2\}$ is inserted to $M$. Since in the second branch, we next apply for each $x\in \{x_1,x_2\}$ a reduction rule in a manner than decreases the measure by at least $\frac{1}{4}$ (because $x\in M$, else the measure does not necessarily decrease), the branching vector is at least as good as $(1\frac{1}{4},1\frac{1}{4})$, whose root is smaller than~$3.188^{0.5}$.}

\begin{branchrule}\label{branch:17-}{\normalfont
[There is $v\in M$ s.t.~$\{u\}=\N(v)\setminus V_T$ and $|X|=2$, where $X=\N(u)\setminus V_T$. Let $x$ be the vertex in $X$ such that $\paths(M\cup N,x,V\setminus(V_T\cup F\cup\{u\}))\neq\emptyset$ and $|Y|\geq 1$, where $Y=\N(x)\setminus(V_T\cup\{u\})$.\footnote{There is such a vertex in $X$, since otherwise Rule \ref{red:6}, \ref{red:reach} or \ref{branch:Mreduce} was applied. Moreover, there is exactly one such vertex in $X$, since otherwise Rule \ref{rule:M2b} was applied.} Then, $|Y|\geq 2$.] Let $\widetilde{Y}=\N(x)\setminus(\internal(T)\cup\{u\})$.
\begin{enumerate}
\item If \alg{Alg}$(T,L\cup\{v\},M\setminus\{v\},F)$ accepts: Accept.
\item Else if \alg{Alg}$(T'=(V_T\cup\{u\}\cup X, E_T\cup\{(v,u)\}\cup\{(u,w): w\in X\}),L\cup\{x\},(M\cup X)\setminus\{v,x\},F\cup \widetilde{Y})$ accepts: Accept.
\item Return \alg{Alg}$(T'=(V_T\cup\{u\}\cup X, E_T\cup\{(v,u)\}\cup\{(u,w): w\in X\}\cup\{(x,w): w\in Y\}),L,(M\cup X\cup Y)\setminus(\{v,x\}\cup F),F)$ accepts: Accept.
\end{enumerate}
}\end{branchrule}

{\noindent This rule is exaustive in the sense that we either determine that $v$ is a leaf (branch 1), or an internal vertex (branches 2 and 3), where in the latter case, we continue and determine whether $x$ is a leaf (branch 2) or an internal vertex (branch 3). As in previous rules, upon determining that a vertex is an internal vertex, we insert its neighbors outside the tree as its children. In the second branch, we can safely insert $\widetilde{Y}$ to $F$, since otherwise, if there is a solution to the instance in this branch excluding the requirement that $\widetilde{Y}$ is inserted to $F$, we can construct a solution to the instance in the first branch (which contradicts the fact that \alg{Alg} rejected it). The dependency claim is preserved since $X\setminus\{x\}$ is inserted to $M$, where in the third branch, $Y\setminus F$ is also inserted to $M$. Observe that for the vertex in $X\setminus\{x\}$, since it is inserted to $M$ and Rule \ref{rule:M2b} was not applied, we next apply a reduction rule where the measure is decreased by at least $\frac{1}{4}$. Moreover $(X\setminus\{x\})\cap\widetilde{Y}=\emptyset$ and $\widetilde{Y}\setminus F\neq\emptyset$ (since $\paths(M\cup N,x,V\setminus(V_T\cup F\cup\{u\}))\neq\emptyset$ and Rule \ref{rule:M2b} was not applied). Therefore, the branching vector is at least as good as $(1\frac{1}{4},1\frac{1}{4}+(1+|\widetilde{Y}\setminus F|,(|Y|-1)-\frac{1}{4}|Y\setminus F|))$. The worst case is obtained when $\widetilde{Y}\subseteq Y$, $|Y|=2$ and $|Y\setminus F|=1$; thus, the branching vector is at least as good as $(1\frac{1}{4},1\frac{1}{4}+(2,\frac{3}{4}))=(1\frac{1}{4},3\frac{1}{4},2)$, whose root is smaller than $3.188^{0.5}$.}

\begin{figure}[!ht]\centering
\frame{\includegraphics[scale=0.6]{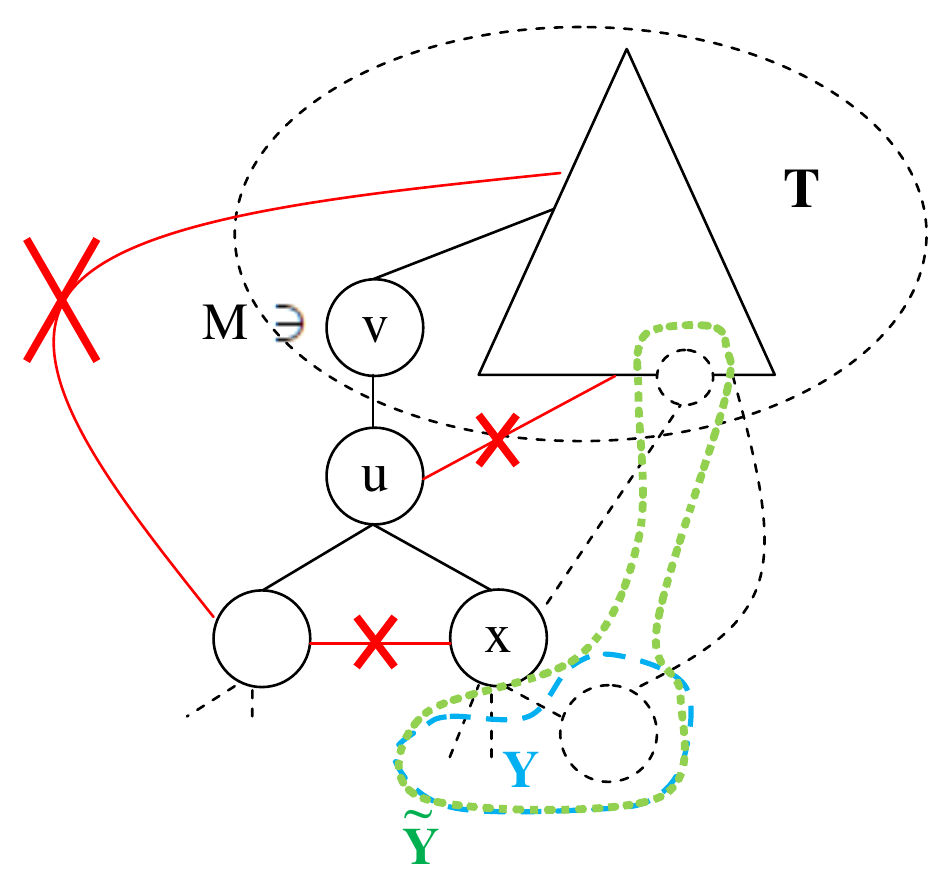}}
\caption{The case handled by Rule \ref{branch:17-}.}
\end{figure}

\begin{branchrule}\label{branch:17--}{\normalfont
[There is $v\in M$ such that $\{u\}=\N(v)\setminus V_T$ and $|X|=2$, where $X=\N(u)\setminus V_T$. Let $x$ be the vertex in $X$ such that $\paths(MN,x,V\setminus(V_T\cup F\cup\{u\}))\neq\emptyset$ and $\{y\}=\N(x)\setminus(V_T\cup\{u\})$.\footnote{There is exactly one such vertex in $X$, since otherwise Rule \ref{branch:17-} was applied.} Also, let $Z=\N(y)\setminus(V_T\cup\{x\})$. Then, $|Z|\geq 3$ or there is no $z\in Z$ such that $\paths(M\cup N,z,V\setminus(V_T\cup F\cup\{u,x,y\}))=\emptyset$.] Let $\widetilde{Z}=Z\setminus F$ if $|Z|\geq 3$, and otherwise $\widetilde{Z}=\emptyset$.
\begin{enumerate}
\item If \alg{Alg}$(T,L\cup\{v\},M\setminus\{v\},F)$ accepts: Accept.
\item Else if \alg{Alg}$(T'=(V_T\cup\{u\}\cup X, E_T\cup\{(v,u)\}\cup\{(u,w): w\in X\}),L\cup\{x\},(M\cup X)\setminus\{v,x\}\,F\cup \{y\})$ accepts: Accept.
\item Return \alg{Alg}$(T'=(V_T\cup\{u,y\}\cup X\cup Z, E_T\cup\{(v,u),(x,y)\}\cup\{(u,w): w\in X\})\cup\{(y,w): w\in Z\},L,(M\cup X\cup\widetilde{Z})\setminus\{v,x\},F)$ accepts: Accept.
\end{enumerate}
}\end{branchrule}

{\noindent The correctness of this rule is similar to the previous one, except that now once we determine that $x$ is an internal vertex (in branch 3), we also determine that $y$ is an internal vertex (this follows as argued for the insertion of $u$ as internal vertex once we determine that $v$ is an internal vertex; observe that $|Z|\geq 2$, since otherwise Rule \ref{branch:Mreduce} was applied). Moreover, if $|Z|=2$, we do not need to insert $Z$ to $M$ since then, by the condition of this rule, there is no $z\in Z$ such that $\paths(M\cup N,z,V\setminus(V_T\cup F\cup\{u,x,y\}))=\emptyset$. As in the previous rule, though noting that now $(X\setminus\{x\})\cup\widetilde{Z}$ is inserted to $M$ in the third branch, the branching vector is at least as good as $(1\frac{1}{4},1\frac{1}{4}+(2,(|Z|-1)-\frac{1}{4}|\widetilde{Z}|))$, which is at least as good as $(1\frac{1}{4},3\frac{1}{4},2\frac{1}{4})$, whose root is smaller than $3.188^{0.5}$.}

\begin{figure}[!ht]\centering
\frame{\includegraphics[scale=0.55]{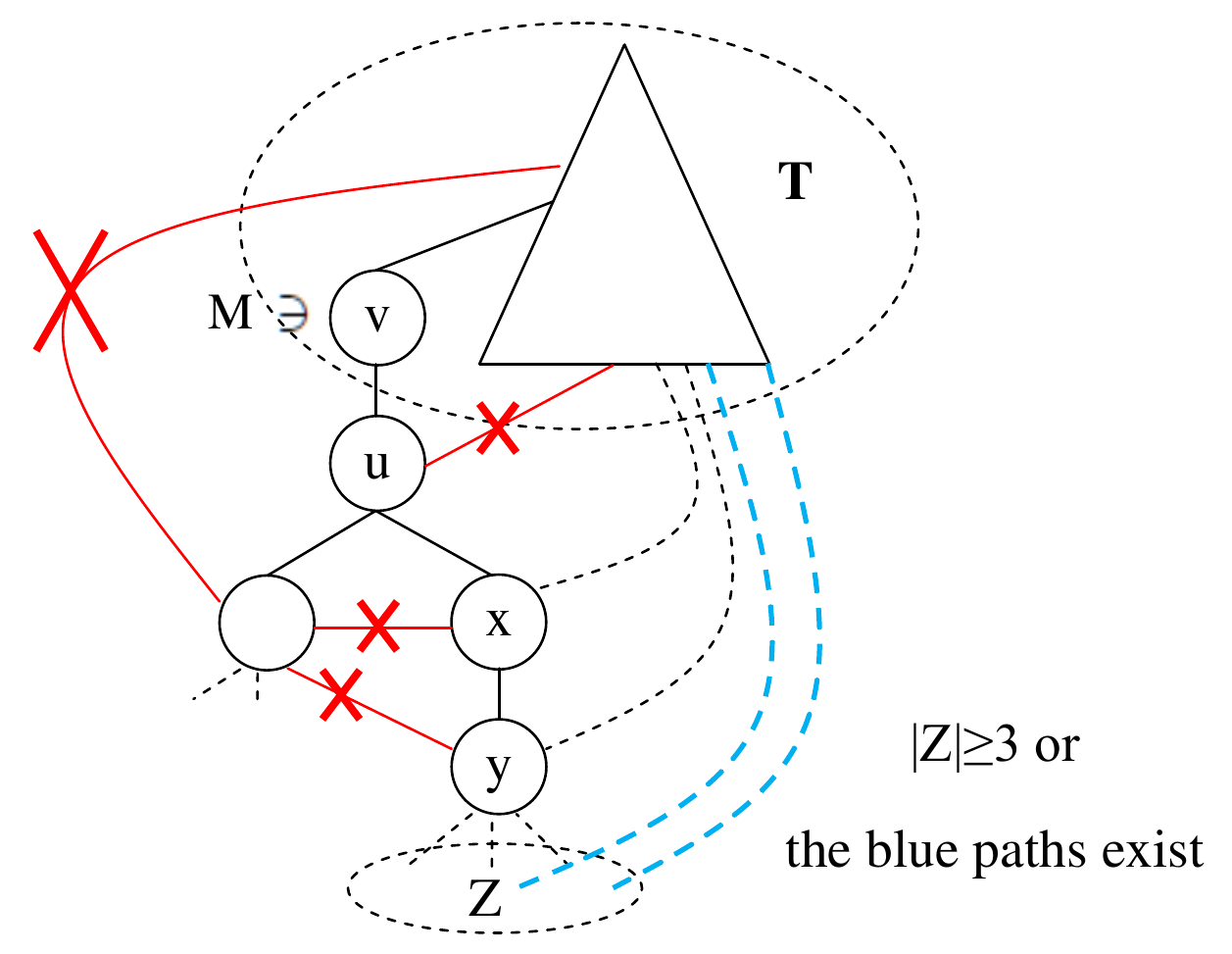}}
\caption{The case handled by Rule \ref{branch:17--}.}
\end{figure}

\begin{branchrule}\label{red:Mend}{\normalfont
[There is $v\in M$ such that $\{u\}=\N(v)\setminus V_T$ and $|X|=2$, where $X=\N(u)\setminus V_T$.] Let $x$ be the vertex in $X$ such that $\paths(M\cup N,x,V\setminus(V_T\cup F\cup\{u\}))\neq\emptyset$ and $\{y\}=\N(x)\setminus(V_T\cup\{u\})$. Also, let $Z=\N(y)\setminus(V_T\cup\{x\})$.
\begin{enumerate}
\item If \alg{Alg}$(T,L\cup\{v\},M\setminus\{v\},F)$ accepts: Accept.
\item Else if \alg{Alg}$(T'=(V_T\cup\{u\}\cup X, E_T\cup\{(v,u)\}\cup\{(u,w): w\in X\}),L\cup\{x\},(M\cup X)\setminus\{v,x\},F\cup\{y\})$ accepts: Accept.
\item Return \alg{Alg}$(T'=(V_T\cup\{u\}\cup X, E_T\cup\{(v,u),(x,y)\}\cup\{(u,w): w\in X\})\cup\{(y,w): w\in Z\},L,(M\cup X\cup Z)\setminus\{v,x\},F)$ accepts: Accept.
\end{enumerate}
}\end{branchrule}

{\noindent The correctness of this rule is similar to the previous one. The correctness of the dependency claim is preserved since now, in the third branch, we insert $Z$ to $M$. Observe that since the previous rule was not applied, $|Z|=2$ and there is a vertex $z\in Z$ such that $\paths(M\cup N,z,V\setminus(V_T\cup F\cup\{u,x,y\}))=\emptyset$; then, in the third branch we next apply a reduction rule that decreases the measure by at least $\frac{1}{4}$, Rule 10 or 11, where Rule 11 is applied with a branching vector at least as good as $(1\frac{1}{4},2\frac{1}{4})$. Thus, the branching vector is at least as good as $(1\frac{1}{4},1\frac{1}{4}+(2,(|Z|-1)-\frac{1}{4}|Z|+\frac{1}{4}))=(1\frac{1}{4},3\frac{1}{4},2)$, whose root is smaller than~$3.188^{0.5}$.}

\bigskip
{\noindent\bf For now on, since Rules \ref{red:deg2deg2}--\ref{red:Mend} were not applied, $M=\emptyset$.}

\subsection{Vertices in $N$ Without a Sibling in $N$}

\begin{branchrule}\label{branch:xsN3}{\normalfont
[There is $v\!\in\! N$ s.t.~$\siblings(v)\!\cap\! N\!=\!\emptyset$ and $|X|\!\geq\! 3$, where $X\!=\!\N(v)\!\setminus\! V_T$] 
\begin{enumerate}
\item If \alg{Alg}$(T,L\cup\{v\},M,F\cup X)$ accepts: Accept.
\item Return \alg{Alg}$(T'=(V_T\cup X, E_T\cup\{(v,u): u\in X\}),L,M\cup (X\setminus F),F)$.
\end{enumerate}
}\end{branchrule}

{\noindent The rule is exhaustive in the sense that we determine that either $v$ is a leaf or $v$ is an internal vertex (where in the latter case, we have already established that we can connect the neighbors of $v$ outside $T$ as children of $v$---see Rule \ref{red:reach}). Thus, it only remains to show that in the first branch, we can insert $X$ to $F$. Recall that by the dependency claim, $|\siblings(v)|\leq 1$. First, suppose that either $\siblings(v)=\emptyset$ or there exists $s\in\siblings(v)$ such that $s\in L$. Then, it is clear that we can insert $X$ to $F$ by considering the explanation given for the first branch of Rule \ref{branch:worse} (see Section \ref{sec:examples}): if there is a solution where $v$ is a leaf and some vertex in $X$ is not a leaf, we can disconnect $v$ from $\parent(v)$ and reattach it to this vertex, disconnect the sibling of $v$ (if one exists) from $\parent(v)$ and reattach it to another vertex in $V\setminus(\internal(\widetilde{T})\cup L'\cup F')$, overall obtaining a solution that contradicts the correctness of the dependency claim (in particular, $\parent(v)$ is a leaf in this solution). 

If the supposition is not true, then by the dependency claim, we have that there exists $s\in\siblings(v)$ such that $s\in \children_1(T)$. Then, let $S$ be a solution to $(T,L\cup\{v\},M,F)$. Assume that there is a vertex $w\in X$ that is not a leaf in $S$. We start by disconnecting the leaf $v$ from its parent and attaching it to $w$, obtaining a solution $S'$ with the same number of leaves as $S$. Next, we disconnect $s$ from $\parent(T)$ and reattach a vertex $q$ in its subtree to a vertex $p\in V\setminus(\internal(\widetilde{T})\cup L'\cup F')$ (whose existence is guaranteed by the dependency claim). We thus obtain a spanning tree $S''$, where all the leaves in $S'$ are leaves in $S''$, excluding possibly $q$ and $p$. The vertex $\parent(v)$ is a new leaf in $S''$, and if $p$ was a leaf in $S'$, then $s$ is a new leaf in $S''$. We have that $S''$ has at least as many leaves as $S$. Thus, we obtain a solution, $S''$, where the parent of $v$ and $s$ in $T$ is a leaf. By our construction, $S''$ complies with $(\widetilde{T},L'\cup F')$. This contradicts the dependency claim. Therefore, we showed that in the first branch, it is safe to insert $X$ to $F$.

Since in the second branch the vertices in $X\setminus F$ are inserted to $M$, the correctness of the dependency claim is preserved. The branching vector is $(1+|X\setminus F|,(|X|-1)-\frac{1}{4}|X\setminus F|)$. At worst, $|X|=3$ and $X\subseteq F$, which results in the branching vector $(1,2)$, whose root is smaller than $3.188^{0.5}$.}

\begin{figure}[!ht]\centering
\frame{\includegraphics[scale=0.5]{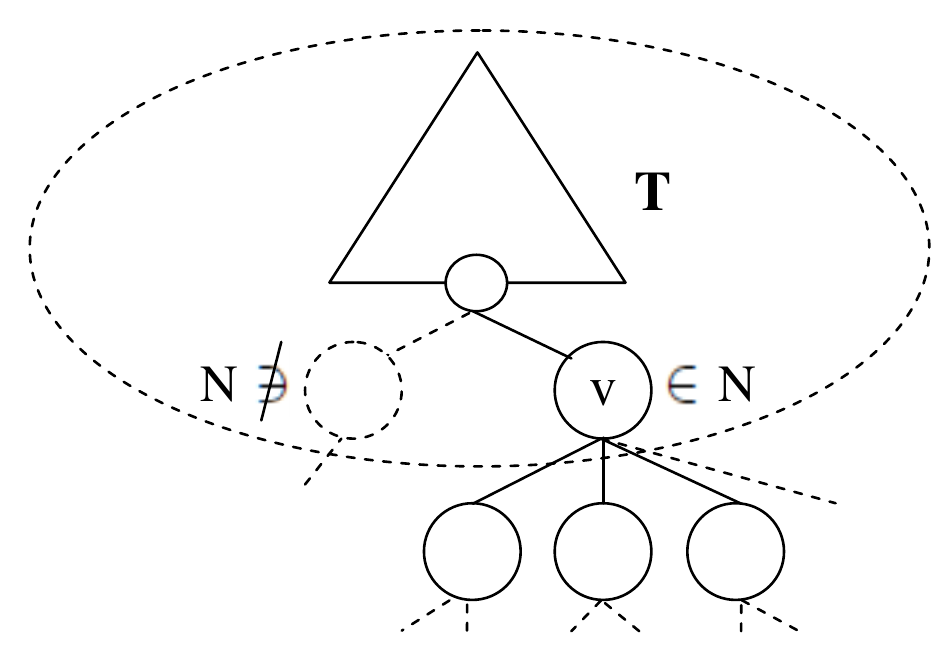}}
\caption{The case handled by Rule \ref{branch:xsN3}.}
\end{figure}

\begin{branchrule}\label{branch:17}{\normalfont
[There is $v\!\in\! N$ s.t.~$\siblings(v)\!\cap\! N\!=\!\emptyset$ and $|X|\!=\!2$, where $X\!=\!\N(v)\!\setminus\! V_T$]
\begin{enumerate}
\item If \alg{Alg}$(T,L\cup\{v\},M,F\cup X)$ accepts: Accept.
\item Return \alg{Alg}$(T'=(V_T\cup X, E_T\cup\{(v,u): u\in X\}),L,M,F)$.
\end{enumerate}
}\end{branchrule}

{\noindent Again, this rule is exhaustive in the sense that we determine that either $v$ is a leaf or $v$ is an internal vertex (in which case we can connect the neighbors of $v$ outside $T$ as children of $v$). In the first branch, as in the first branch of Rule \ref{branch:xsN3}, we can insert $X$ to $F$. The dependency claim holds in the second branch since \alg{Alg} rejected the instance in the first branch and Rule \ref{red:reach} was not applied. Since Rule \ref{branch:neisinF} was not applied, $X\setminus F\neq\emptyset$. Thus, the branching vector, $(1+|X\setminus F|,1)$, is at least as good as $(2,1)$, whose root is smaller than $3.188^{0.5}$.}

\begin{figure}[!ht]\centering
\frame{\includegraphics[scale=0.5]{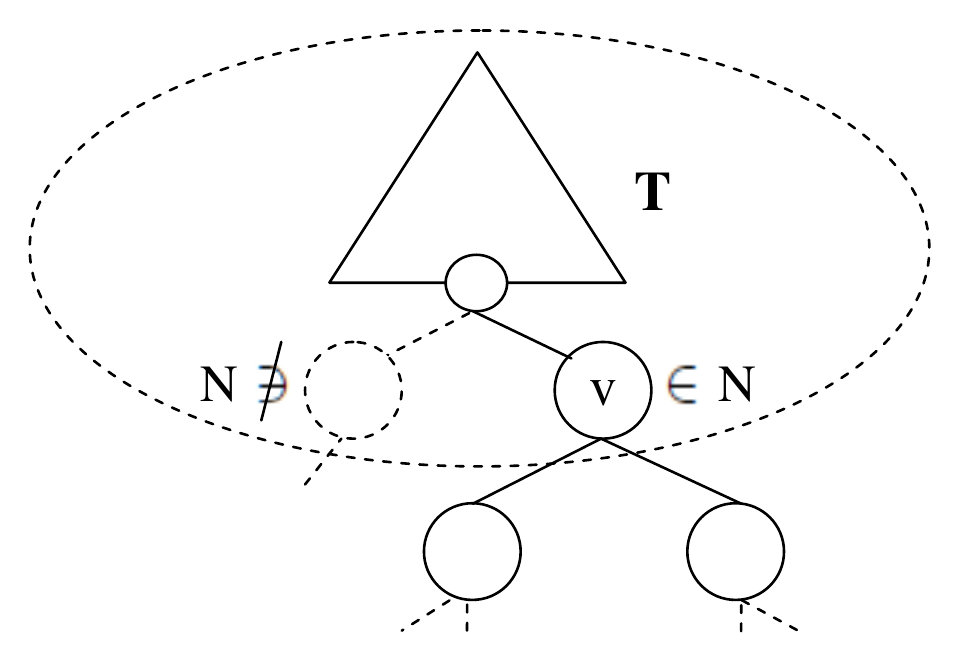}}
\caption{The case handled by Rule \ref{branch:17}.}
\end{figure}

{\noindent Next, note that since previous reduction rules were note applied, if $v\in N$ such that $\siblings(v)\cap N=\emptyset$ and $\{u\}=N(v)\setminus V_T$, then $u$ is outside $F$ and has at least two neibors outside~the~tree.}

\begin{branchrule}\label{branch:18}{\normalfont
[There is $v\in N$ such that $\siblings(v)\cap N=\emptyset$, $\{u\}=\N(v)\setminus V_T$ and $|X|\geq 3$, where $X=\N(u)\setminus V_T$]
\begin{enumerate}
\item If \alg{Alg}$(T,L\cup\{v\},M,F\cup \{u\})$ accepts: Accept.
\item Return \alg{Alg}$(T'=(V_T\cup\{u\}\cup X, E_T\cup\{(v,u)\}\cup\{(u,w): w\in X\}),L,M\cup X,F)$.
\end{enumerate}
}\end{branchrule}

{\noindent Again, this rule is exhaustive in the sense that we determine that either $v$ is a leaf (in which case we can insert $u$ to $F$) or $v$ is an internal vertex (in which case we can connect the neighbors of $u$ outside $T$ as children of $u$). Thus, it remains to argue that in the second branch, where $v$ is an internal vertex, we can also determine that $u$ is an internal vertex. This follows from the fact that if there is a solution to $(T'=(V_T\cup\{u\}, E_T\cup\{(v,u)\}),L\cup\{u\},M,F)$, we can disconnect $u$ from $v$ and attach it to some other neighbor in $V\setminus(\internal(T)\cup L\cup F)$ (this is possible, else Rule \ref{red:reach} was applied), and obtain a solution for the instance in the first branch---a contradiction. The dependency claim holds in the second branch since $X$ is inserted to $M$. The branching vector, $(2,(|X|-1)-\frac{1}{4}|X|)=(2,\frac{3}{4}|X|-1)$, is at least as good as $(2,\frac{5}{4})$, whose root is smaller than $3.188^{0.5}$.}

\begin{figure}[!ht]\centering
\frame{\includegraphics[scale=0.5]{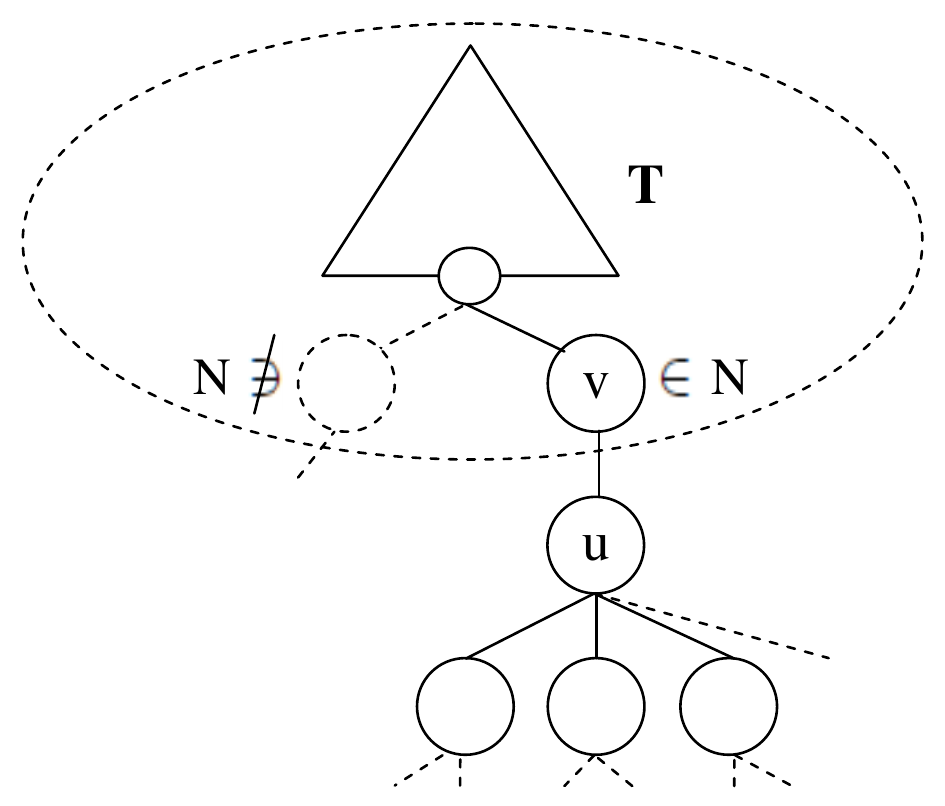}}
\caption{The case handled by Rule \ref{branch:18}.}
\end{figure}

\begin{branchrule}\label{branch:19}{\normalfont
[There is $v\in N$ s.t.~$\siblings(v)\cap N=\emptyset$, $\{u\}=\N(v)\setminus V_T$, and (for all $x\in X$, $\paths(N,x,V\setminus(V_T\cup F\cup \{u\}))\neq\emptyset$), where $X=\N(u)\setminus V_T$]
\begin{enumerate}
\item If \alg{Alg}$(T,L\cup\{v\},M,F\cup\{u\})$ accepts: Accept.
\item Return \alg{Alg}$(T'=(V_T\cup\{u\}\cup X, E_T\cup\{(v,u)\}\cup\{(u,w): w\in X\}),L,M,F)$.
\end{enumerate}
}\end{branchrule}

{\noindent The correctness follows similarly as in the previous rule. Since the previous rule was not applied, $|X|=2$, and by the condition of the rule, (for all $x\in X$, $\paths(N,x,V\setminus(V_T\cup F\cup \{u\}))\neq\emptyset$); therefore, by the order of the branches, the dependency claim is preserved. The branching vector, $(2,1)$, has a root smaller than $3.188^{0.5}$.}

\begin{figure}[!ht]\centering
\frame{\includegraphics[scale=0.5]{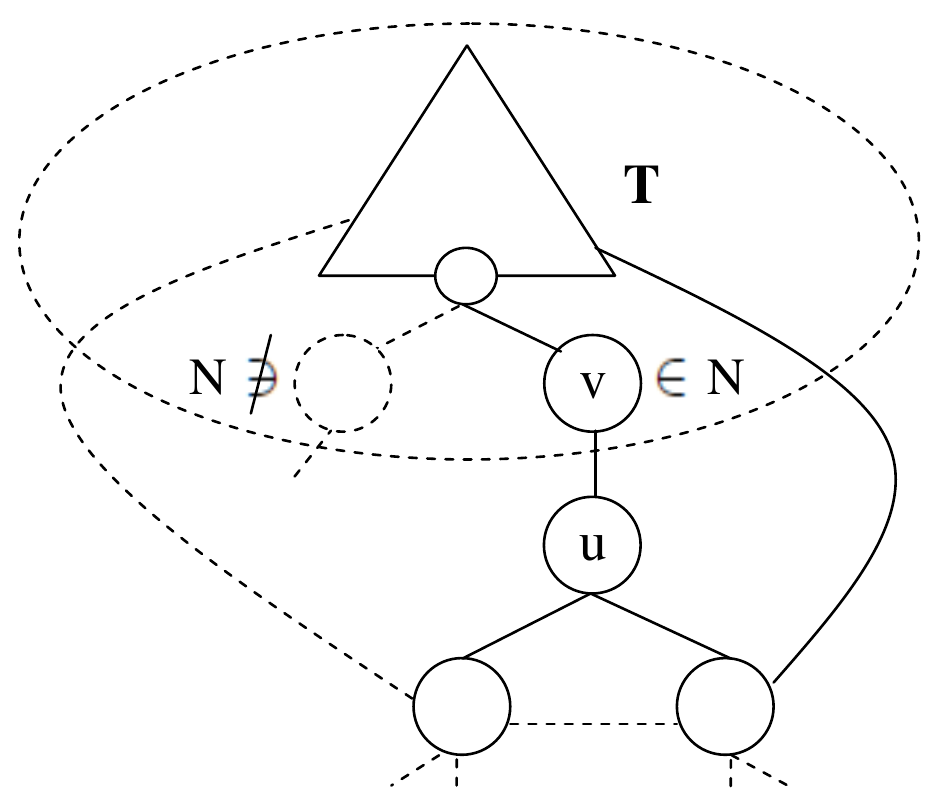}}
\caption{The case handled by Rule \ref{branch:19}.}
\end{figure}

\begin{reducerule}\label{red:20}{\normalfont 
[There is $v\!\in\! N$ s.t.~$\siblings(v)\!\cap\! N\!=\!\emptyset$ and $\{u\}\!=\!\N(v)\!\setminus\! V_T$] Let $X\!=\!\N(u)\!\setminus\! V_T$.
Return \alg{Alg}$(T'=(V_T\cup\{u\}\cup X, E_T\cup\{(v,u)\}\cup\{(u,w): w\in X\}),L,M\cup X,F)$.
}\end{reducerule}

{\noindent Since the two previous rules were not applied, $|X|=2$, and there is $x\in X$ such that $\paths(N,x,V\setminus(V_T\cup F\cup \{u\}))=\emptyset$; therefore, if there is a solution, it contains $u$ as an internal vertex. Thus, if a solution contains $v$ as a leaf, we can disconnect $v$ and reattach it to $u$, while also disconnecting the sibling of $v$ (if one exists) and reattaching its subtree to a vertex in $V\setminus(\internal(\widetilde{T})\cup L'\cup F')$, obtaining a solution that contradicts the dependency claim (as in Rule \ref{branch:xsN3}).
We can therefore safely determine that $v$ and $u$ are internal vertices. The measure decreases by $\frac{1}{2}$ (since $u$ is inserted to $\children_2(T)$ and $X$ is inserted to $M$).}

\bigskip
{\bf\noindent Overall, from now on, $M=\emptyset$, and for any $v\in N$, we have that $|\siblings(v)\cap N|=1$. Also, any vertex in $N$ with exactly two neighbors outside $T$, has a neighbor outside $F$.}

\subsection{Vertices in $N$ with Siblings (in $N$) of One Neighbor Outside~$T$}

For illustrations, follow the figures given in the previous subsection, noting that now the sibling of $v$ belongs to $N$ and has exactly one neighbor outside $T$.

\begin{branchrule}{\normalfont
[There are $v\in N$ and $s\in\siblings(v)\cap N$ such that $|\N(s)\setminus V_T|=1$ and $|X|\geq 3$, where $X=\N(v)\setminus V_T$]
\begin{enumerate}
\item If \alg{Alg}$(T,L\cup\{v\},M,F\cup X)$ accepts: Accept.
\item Return \alg{Alg}$(T'=(V_T\cup X, E_T\cup\{(v,u): u\in X\}),L,M\cup (X\setminus F)\cup\{s\},F)$.
\end{enumerate}
}\end{branchrule}

{\noindent The correctness follows by arguments similar to those in Rule \ref{branch:xsN3}. Similarly, noting that $s$ is inserted to $M$ (in the second branch), the preservation of the dependency claim follows. The branching vector is $(1+|X\setminus F|,(|X|-1)-\frac{1}{4}|X\setminus F|-\frac{1}{4})$. At worse, $|X|=3$ and $X\subseteq F$. Thus, the vector is at least as good as $(1,1\frac{3}{4})$, whose root is smaller than $3.188^{0.5}$.}

\begin{branchrule}{\normalfont
[There are $v\in N$ and $s\in\siblings(v)\cap N$ such that $|\N(s)\setminus V_T|=1$ and $|X|=2$, where $X=\N(v)\setminus V_T$]
\begin{enumerate}
\item If \alg{Alg}$(T,L\cup\{v\},M,F\cup X)$ accepts: Accept.
\item Return \alg{Alg}$(T'=(V_T\cup X, E_T\cup\{(v,u): u\in X\}),L,M\cup\{s\},F)$.
\end{enumerate}
}\end{branchrule}

{\noindent The correctness follows by arguments similar to those in Rule \ref{branch:17}. Similarly, noting that $s$ is inserted to $M$, the preservation of the dependency claim follows. The branching vector is $(1+|X\setminus F|,1-\frac{1}{4})$, which is at least as good as $(2,\frac{3}{4})$, whose root is smaller than $3.188^{0.5}$.}

\begin{branchrule}{\normalfont 
[There are $v\in N$ and $s\in\siblings(v)\cap N$ such that $|\N(s)\setminus V_T|=1$, $\{u\}=\N(v)\setminus V_T$ and $|X|\geq 3$, where $X=\N(u)\setminus V_T$]
\begin{enumerate}
\item If \alg{Alg}$(T,L\cup\{v\},M,F\cup\{u\})$ accepts: Accept.
\item Return \alg{Alg}$(T'=(V_T\cup\{u\}\cup X, E_T\cup\{(v,u)\}\cup\{(u,w): w\in X\}),L,M\cup X\cup\{s\},F)$.
\end{enumerate}
}\end{branchrule}

{\noindent The correctness follows by arguments similar to those in Rule \ref{branch:18}. Similarly, noting that $s$ is inserted to $M$, the preservation of the dependency claim follows. The branching vector is $(2,(|X|-1)-\frac{1}{4}|X|-\frac{1}{4})=(2,\frac{3}{4}|X|-1\frac{1}{4})$, which is at least as good as $(2,1)$, whose root is smaller than $3.188^{0.5}$.}

\begin{branchrule}{\normalfont
[There is $v\in N$ and $s\in\siblings(v)\cap N$ such that $|\N(s)\setminus V_T|=1$, $\{u\}=\N(v)\setminus V_T$ and (for all $x\in X$, $\paths(N,x,V\setminus(V_T\cup F\cup \{u\}))\neq\emptyset$), where $X=\N(u)\setminus V_T$]
\begin{enumerate}
\item If \alg{Alg}$(T,L\cup\{v\},M,F\cup\{u\})$ accepts: Accept.
\item Return \alg{Alg}$(T'=(V_T\cup\{u\}\cup X, E_T\cup\{(v,u)\}\cup\{(u,w): w\in X\}),L,M\cup\{s\},F)$.
\end{enumerate}
}\end{branchrule}

{\noindent The correctness follows by arguments similar to those in Rule \ref{branch:19}. Similarly, noting that $s$ is inserted to $M$, the preservation of the dependency claim follows. The branching vector is $(2,1-\frac{1}{4})=(2,\frac{3}{4})$, whose root is smaller than $3.188^{0.5}$.}

\begin{reducerule}\label{red:1sibend}{\normalfont 
[There are $v\in N$ and $s\in\siblings(v)\cap N$ such that $|\N(s)\setminus V_T|=1$ and $\{u\}=\N(v)\setminus V_T$] Let $X=\N(u)\setminus V_T$.
Return \alg{Alg}$(T'=(V_T\cup\{u\}\cup X, E_T\cup\{(v,u)\}\cup\{(u,w): w\in X\}),L,M\cup X\cup \{s\},F)$.
}\end{reducerule}

{\noindent The correctness follows by arguments similar to those in Rule \ref{red:20}. Similarly, noting that $s$ is inserted to $M$, the preservation of the dependency claim follows. The measure decreases by at least $(1-\frac{3}{4})=\frac{1}{4}$.}

\bigskip
{\bf\noindent For now on, if $v\in N$, both $v$ and its sibling have (each) at least two neighbors outside the tree (if exactly two, not both in $F$).}

\subsection{Siblings in $N$ Having a Vertex Reachable Only from Them}

\begin{branchrule}\label{rule:29*}{\normalfont
[There are $v\in N$, $s\in\siblings(v)\cap N$ and $u\in V\setminus V_T$ such that $\paths(N\setminus\{v,s\},u,V\setminus(V_T\cup F))=\emptyset$ and $|X|\geq 3$, where $X=\N(v)\setminus V_T$]
Let $Y=\N(s)\setminus V_T$. 
\begin{enumerate}
\item If \alg{Alg}$(T'=(V_T\cup Y, E_T\cup\{(s,u): u\in Y\}),L\cup\{v\},M\cup Y,F)$ accepts: Accept.
\item Return \alg{Alg}$(T'=(V_T\cup X, E_T\cup\{(v,u): u\in X\}),L,M\cup X\cup\{s\},F)$.
\end{enumerate}
}\end{branchrule}

{\noindent This rule is exhaustive in the sense that we determine that either $v$ is a leaf or an internal vertex, where upon determining that a vertex is an internal vertex, we insert its neighbors outside $T$ as its children. In the first branch we also determine that $s$ is an internal vertex, since otherwise we cannot reach the vertex $u$. Clearly, the dependency claim holds (since in the first branch, we insert $Y$ to $M$, and in the second branch, we insert $X\cup\{s\}$ to $M$). The branching vector is $(1+(|Y|-1)-\frac{1}{4}|Y|, (|X|-1)-\frac{1}{4}(|X|+1))=(\frac{3}{4}|Y|,\frac{3}{4}|X|-\frac{5}{4})$, which is at least as good as $(1.5,1)$ whose root is smaller than $3.188^{0.5}$.}

\begin{figure}[!ht]\centering
\frame{\includegraphics[scale=0.5]{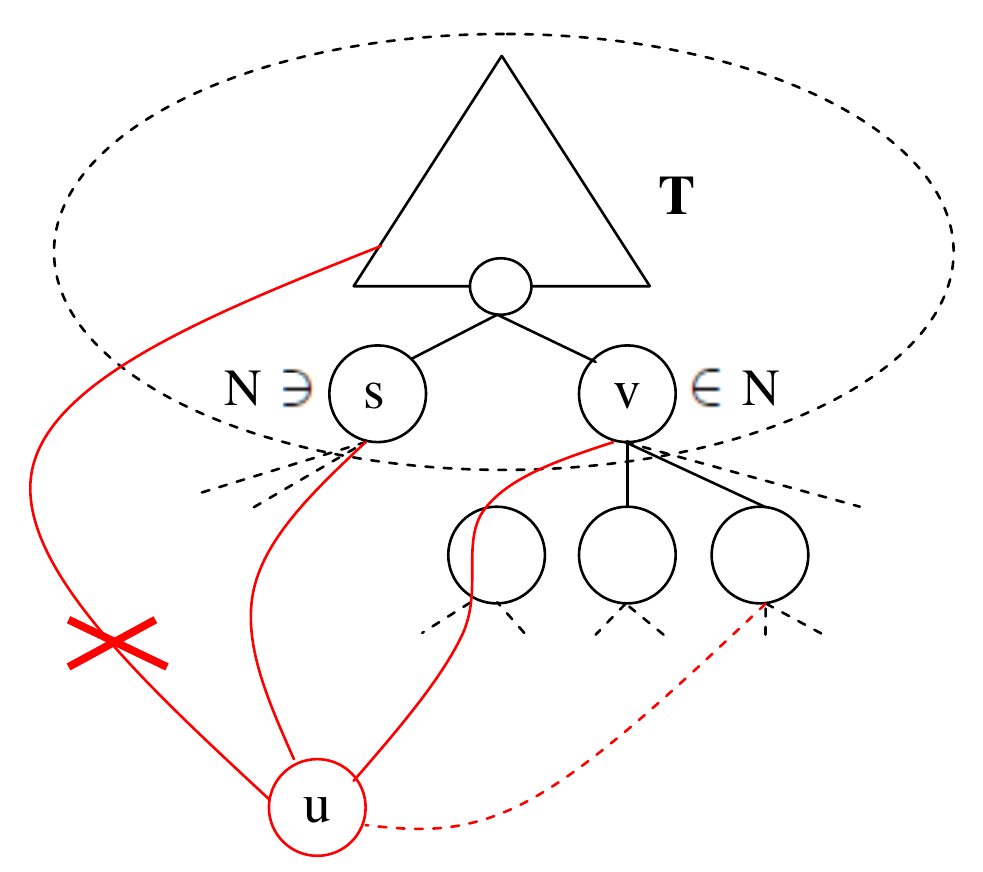}}
\caption{The case handled by Rule \ref{rule:29*}.}
\end{figure}

\begin{branchrule}\label{branch:Mafter}{\normalfont
[There are $v\in N$, $s\in\siblings(v)\cap N$ and $u\in V\setminus V_T$ such that $\paths(N\setminus\{v,s\},u,V\setminus(V_T\cup F))=\emptyset$]
Let $X=\N(v)\setminus V_T$ and $Y=\N(s)\setminus V_T$. 
\begin{enumerate}
\item If \alg{Alg}$(T'=(V_T\cup Y, E_T\cup\{(s,u): u\in Y\}),L\cup\{v\},M\cup Y,F)$ accepts: Accept.
\item Return \alg{Alg}$(T'=(V_T\cup X, E_T\cup\{(v,u): u\in X\}),L,M\cup\{s\},F)$.
\end{enumerate}
}\end{branchrule}

{\noindent The correctness and preservation of the dependency claim follow as in Rule \ref{rule:29*} (now, since $|X|=2$, we do not need to insert $X$ to $M$ in the second branch). Since Rule \ref{rule:29*} was not applied, $|X|=|Y|=2$. Now, observe that as long as we have a vertex in $M$, we terminate the execution (by applying Rule \ref{red:stop1}, \ref{red:stop2} or \ref{red:stop3}), or apply a reduction rule where the measure decreases by at least $0.25$ (this statement is true because we have a vertex in $M$, otherwise some reduction rules may not decreases the measure), or apply a branching rule (among Rules \ref{branch:neisinF}--\ref{red:Mend}) whose branching vector is at worst $(1.25,1.25)$ or a combination of $(1.25,1.25)$ and a vector whose root is smaller than $3.188^{0.5}$. In the first branch, we insert two verices, $y_1,y_2\in Y$, to $M$. If for the first one examined among them, $y_1$, we apply a branching rule whose branching vector is $(1.25,1.25)$, then, by the order and conditions of our rules, there is now a vertex (a neighbor of $y_1$) that in the first subbranch, where $y_1$ is determined to be a leaf, is reachable only from $y_2$, and thus we use a reduction rule to determine (in this subbranch) that $y_2$ is an internal vertex---correspondingly, the measure decreases by at least $0.25$. In the second subbranch, at worse, we apply a branching rule whose branching vector $(1.25,1.25)$. We therefore obtain that at worst, the branching vector of this rule is $(2-0.25-0.25+t,0.75+(1.25,1.25))$, where $t=(1.25+0.25,1.25+(1.25,1.25))$. That is, at worst, the branching vector of this rule is $(3,4,4,2,2)$, whose root is smaller than $3.188^{0.5}$.}

\begin{figure}[!h]\centering
\frame{\includegraphics[scale=0.5]{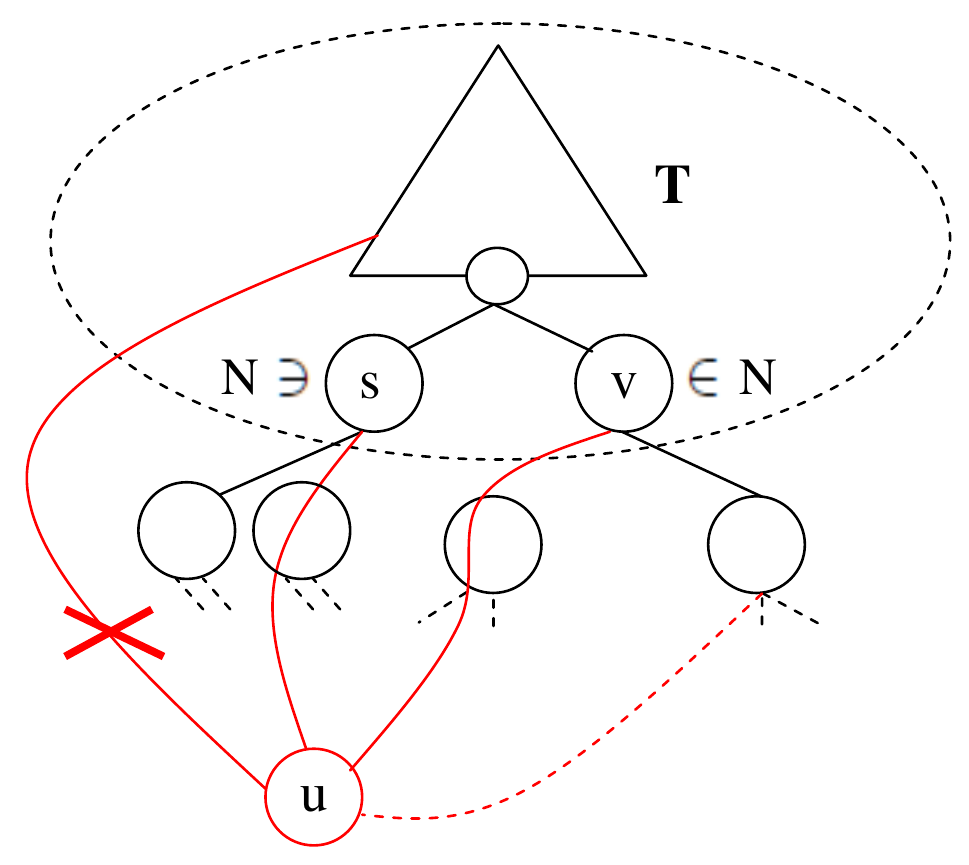}}
\caption{The case handled by Rule \ref{branch:Mafter}.}
\end{figure}

\bigskip
{\bf\noindent For now on, if $v\in N$ and $s\in\siblings(v)\cap N$, there is no $u\in V\setminus V_T$ such that $\paths(N\setminus\{v,s\},u,V\setminus(V_T\cup F))=\emptyset$.}

\subsection{Siblings in $N$ Having a Common Neighbor Outside $T$ and $F$}

\begin{branchrule}\label{rule:31*}{\normalfont
[There are $v\in N$, $s\in\siblings(v)\cap N$ and $u\in (X\cap Y)\setminus F$, where $X=\N(v)\setminus V_T$ and $Y=\N(s)\setminus V_T$]
If $|X|\geq 3$, let $\widetilde{X}=X\setminus F$, and else $\widetilde{X}=\emptyset$. Symmetrically, denote $\widetilde{Y}$. 
\begin{enumerate}
\item If \alg{Alg}$(T'\!=\!(V_T\!\cup\! Y, E_T\!\cup\!\{(s,u)\!: u\!\in\! Y\}),L\!\cup\!\{v\},M\!\cup\! \widetilde{Y},F)$ accepts: Accept.
\item Return \alg{Alg}$(T'=(V_T\cup X, E_T\cup\{(v,u): u\in X\}),L,M\cup \widetilde{X}\cup\{s\},F)$.
\end{enumerate}
}\end{branchrule}

{\noindent This rule is exhaustive in the sense that we determine that either $v$ is a leaf or an internal vertex, where upon determining that a vertex is an internal vertex, we insert its neighbors outside $T$ as its children. We need to argue that in the first branch, it is safe to determine that $s$ is an internal vertex. To this end, it suffices to show that if there is a solution $S$ to $(T,L\cup\{v,s\},M,F)$, then we reach a contradiction. Indeed, we can disconnect the leaves $v$ and $s$, reattaching them to $u$ (in $S$), obtaining a solution with at least as many leaves as $S$ where, in particular, $\parent(v)$ is a leaf---this contradicts the dependency claim. Moreover, since we have already established that there is no $w\in V\setminus V_T$ such that $\paths(N\setminus\{v,s\},w,V\setminus(V_T\cup F))=\emptyset$, and because $s$ is inserted to $M$ in the second branch, the correctness of the dependency claim is preserved. The branching vector is $(1+(|Y|-1)-\frac{1}{4}|\widetilde{Y}|,(|X|-1)-\frac{1}{4}|\widetilde{X}|-\frac{1}{4})$. At worse, $|X|=|Y|=2$, and we obtain the branching vector $(2,0.75)$, whose root is smaller than $3.188^{0.5}$.}

\begin{figure}[!h]\centering
\frame{\includegraphics[scale=0.5]{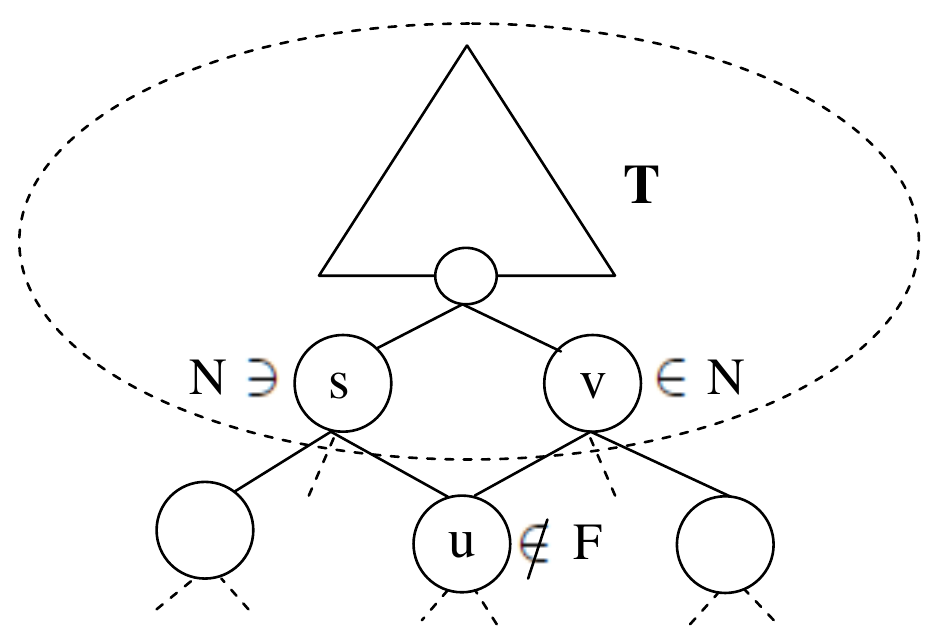}}
\caption{The case handled by Rule \ref{rule:31*}.}
\end{figure}

\bigskip
{\bf\noindent For now on, if $v\in N$ and $s\in\siblings(v)\cap N$, then $(\N(v)\cap\N(s))\setminus V_T\subseteq F$.}

\subsection{Vertices in $N$ with Many Neighbors Outside $T$}

\begin{branchrule}\label{rule:32*}{\normalfont
[There are $v\in N$, $s\in\siblings(v)\cap N$ such that $|X|\geq 3$ and $|X\cap F|\geq 2$, where $X=\N(v)\setminus V_T$]
\begin{enumerate}
\item If \alg{Alg}$(T,L\cup\{v\},M,F)$ accepts: Accept.
\item Return \alg{Alg}$(T'=(V_T\cup X, E_T\cup\{(v,u): u\in X\}),L,M\cup (X\setminus F)\cup\{s\},F)$.
\end{enumerate}
}\end{branchrule}

{\noindent This rule is exhaustive in the sense that we determine that either $v$ is a leaf or an internal vertex, where upon determining that $v$ is an internal vertex, we insert its neighbors outside $T$ as its children. Also, since in the second branch we insert $(X\setminus F)\cup\{s\}$ to $M$, the dependency claim is preserved. The branching vector is $(1,(|X|-1)-\frac{1}{4}|(X\setminus F)\cup\{s\}|)=(1,|X|-\frac{1}{4}|X\setminus F|-\frac{5}{4})$. Since $|X|\geq 3$ and  $|X\cap F|\geq 2$, this branching vector is at least as good as $(1,1.5)$, whose root is smaller than $3.188^{0.5}$.}

\begin{figure}[!h]\centering
\frame{\includegraphics[scale=0.5]{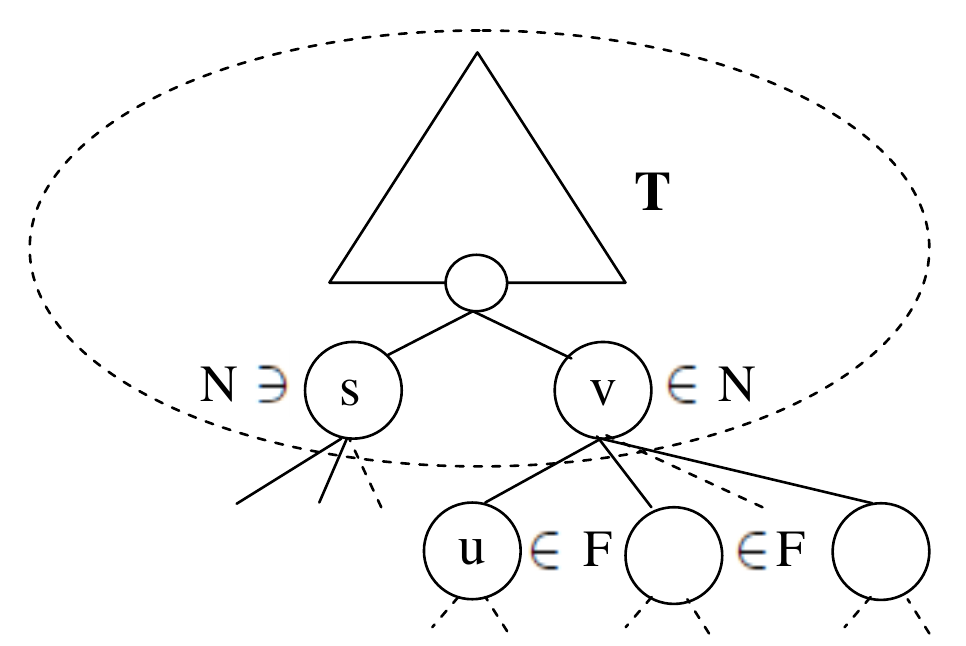}}
\caption{The case handled by Rule \ref{rule:32*}.}
\end{figure}

\begin{branchrule}\label{rule:33*}{\normalfont 
[There are $v\in N$ and $s\in\siblings(v)\cap N$ such that $|X|\geq 3$, where $X=\N(v)\setminus V_T$] Let $Y=\N(s)\setminus V_T$, $Z=X\setminus Y$ and $\widetilde{X}=X\setminus F$. If $|Y|\geq 3$, let $\widetilde{Y}=Y\setminus F$, and else $\widetilde{Y}=\emptyset$. Symmetrically, denote $\widetilde{Z}$. 
\begin{enumerate}
\item If \alg{Alg}$(T,L\cup\{v,s\},M,F\cup X\cup Y)$ accepts: Accept.
\item Else if \alg{Alg}$(T'=(V_T\cup X, E_T\cup\{(v,u): u\in X\}),L\cup\{s\},M\cup\widetilde{X},F)$ accepts: Accept.
\item Else if \alg{Alg}$(T'=(V_T\cup Y, E_T\cup\{(s,u): u\in Y\}),L\cup\{v\},M\cup\widetilde{Y},F)$ accepts: Accept.
\item Return \alg{Alg}$(T'=(V_T\cup X\cup Y, E_T\cup\{(v,u): u\in Z\}\cup\{(s,u): u\in Y\}),L,M\cup\widetilde{Z}\cup\widetilde{Y},F)$.
\end{enumerate}
}\end{branchrule}

{\noindent The rule is exhaustive in the sense that we try all four options to determine the roles of $v$ and $s$, where once a vertex is determined to be an internal vertex, we can attach its neighbors outside $T$ as its children. Also, to see that in the first branch we can insert $X\cup $ to $F$, follow the explanation given for the first branch in Rule \ref{branch:worse}. To obtain a good enough branching vector (explained below), in the fourth branch we attach the common neighbors outside $T$ to $s$ (thus, if $|X|=3$, $|Y|=2$ and there is one common neighbor, in the fourth branch, both $v$ and $s$ have two children---this implies that we do not need to insert their children to $M$). Also, since we have established that there is no $u\in V\setminus V_T$ such that $\paths(N\setminus\{v,s\},u,V\setminus(V_T\cup F))=\emptyset$, the dependency claim is preserved in all branches---observe that once we set a vertex with at least three children, we insert those outside $F$ to $M$. 

The branching vector is $(2+|(X\cup Y)\setminus F|, 1+(|X|-1)-\frac{1}{4}|\widetilde{X}|, 1+(|Y|-1)-\frac{1}{4}|\widetilde{Y}|, (|X\cup Y|-2) -\frac{1}{4}(|\widetilde{Z}|+|\widetilde{Y}|))$. Since the previous rule was not applied, along with rules preceding it (in particular, recall the comments written in the paragraphs starting with ``From now on...''), we have that $|X\setminus F|\geq |X|-1$, $|Y\setminus F|\geq |Y|-1$, $X\cap Y\subseteq F$ and $|Z|\geq 2$ (also note that $|X|\geq 3$). Therefore, we obtain a branching vector that is at least as good as the one of Rule \ref{branch:worse}.}

\begin{figure}[!h]\centering
\frame{\includegraphics[scale=0.5]{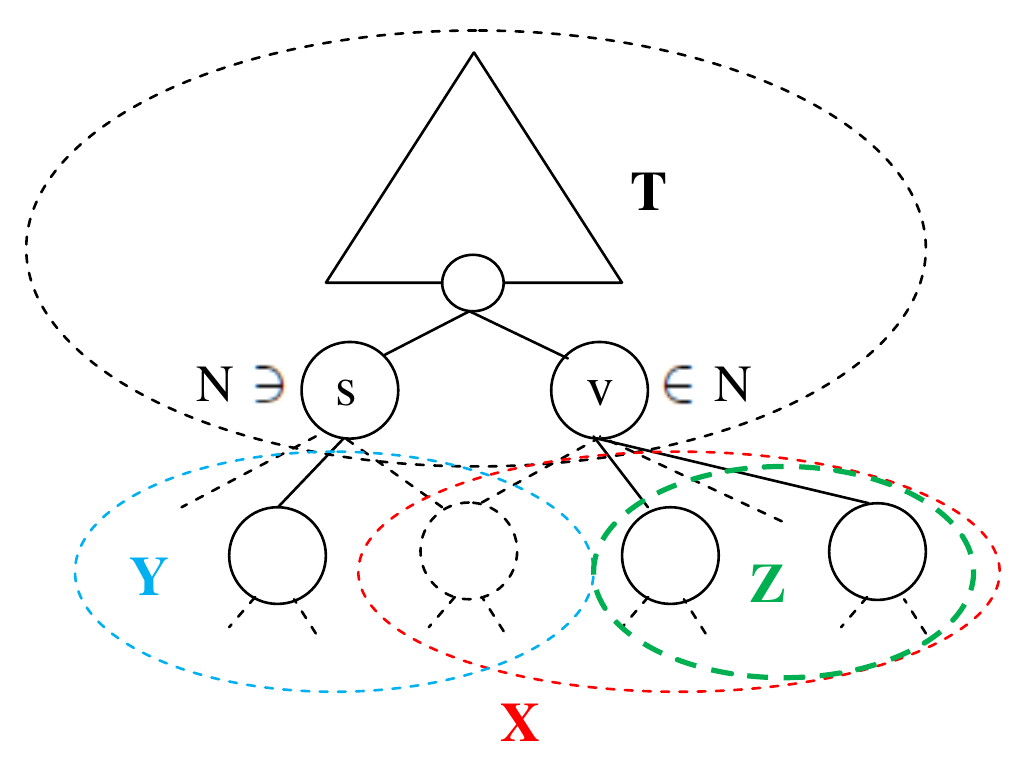}}
\caption{The case handled by Rule \ref{rule:33*}.}
\end{figure}

\bigskip
{\bf\noindent For now on, if $v\in N$, we have that $|\N(v)\setminus V_T|=2$.}

\subsection{Siblings in $N$ with a Common Neighbor (in $F$) Outside $T$}

\begin{branchrule}\label{rule:commonF}{\normalfont 
[There are $v\in N$, $s\in\siblings(v)\cap N$, $u\in F$ and $a,b$ s.t.~$\{a,u\}=\N(v)\setminus V_T$, $\{u,b\}=\N(s)\setminus V_T$, $|X\cup Y|\geq 4$, $|X|\geq 2$ and $|Z|\geq 1$, where $X=\N(a)\setminus(V_T\cup\{v,u,b\})$, $Y=\N(b)\!\setminus(V_T\cup\{s,u,a\})$ and $Z=Y\setminus X$.
Moreover, there is no $w\in Z$ s.t.~$\paths((N\setminus\{v,s\})\cup (X\setminus F),w,V\setminus(V_T\cup F\cup\{v,s,b\}))=\emptyset$.]\\
If $|Z|\geq 3$, let $\widetilde{Z}=Z\setminus F$, and else $\widetilde{Z}=\emptyset$.
\begin{enumerate}
\item If \alg{Alg}$(T,L\cup\{v,s\},M,F\cup \{a,b\})$ accepts: Accept.
\item Else if \alg{Alg}$(T'=(V_T\cup \{a,u\}, E_T\cup\{(v,a),(v,u)\}),L\cup\{s\},M,F)$ accepts: Accept.
\item Else if \alg{Alg}$(T'=(V_T\cup \{u,b\}, E_T\cup\{(s,u),(s,b)\}),L\cup\{v\},M,F)$ accepts: Accept.
\item Return \alg{Alg}$(T'=(V_T\cup\{a,u,b\}\cup X\cup Y, E_T\cup\{(v,a),(v,u),(s,b)\}\cup\{(a,w): w\in X\}\cup\{(b,w): w\in Z\}),L,(M\cup X\cup\widetilde{Z})\setminus F,F)$.
\end{enumerate}
}\end{branchrule}

{\noindent The rule is exhaustive in the sense that we try all four options to determine the roles of $v$ and $s$, where once a vertex is determined to be an internal vertex, we can attach its neighbors outside $T$ as its children. Also, to see that in the first branch we can insert $\{a,b\}$ to $F$, follow the explanation given for the first branch in Rule \ref{branch:worse}. Thus, it is enough to show that in the fourth branch, once determining that $v$ and $s$ are internal vertices, we can also determine that $a$ and $b$ are internal vertices. To this end, suppose that \alg{Alg} did not accept in any of the branches, but there is a solution $S$ to $(T'=(V_T\cup\{a,u,b\},E_T\cup\{(v,a),(v,u),(s,b)\}),L,M,F)$. Since \alg{Alg} did not accept in the fourth branch, at least one among $a$ and $b$ is a leaf in $S$. 
Suppose that $a$ is a leaf. Then, we can disconnect $a$ from $v$ and attach it to another neighbor in $V\setminus V_T$ (the existence of a neighbor as required is gauaranteed since Rule \ref{red:reach}), while disconnecting $u$ and attaching it to $s$, obtaining a solution $S'$ to the instance in third branch---a contradiction. Similarly, if $b$ is a leaf in $S$, we get there is a solution to the instance in the second branch---a contradiction. Therefore, it is safe to determine (in the fourth branch) that $a$ and $b$ are internal vertices.

Since in the fourth branch we insert $(X\setminus F)\cup\widetilde{Z}$ to $M$, and by the condition, there is no $w\!\in\! Y\!\setminus\! X$ s.t.~$\paths((N\setminus\{v,s\})\cup (X\setminus F),w,V\setminus(V_T\cup F\cup\{v,s,b\}))=\emptyset$, the dependency claim is preserved. Now we analyze the branching vector. The branching vector is $(4,2,2,1+(|X\cup Y|-2)-\frac{1}{4}|X\setminus F|-\frac{1}{4}|\widetilde{Z}|)$. Since $|X\cup Y|\geq 4$, $|X|\geq 2$ and $|Z|\geq 1$, this branching vector is at least as good as $(4,2,2,2.25)$, whose root is smaller than $3.188^{0.5}$.}

\begin{figure}[!h]\centering
\frame{\includegraphics[scale=0.5]{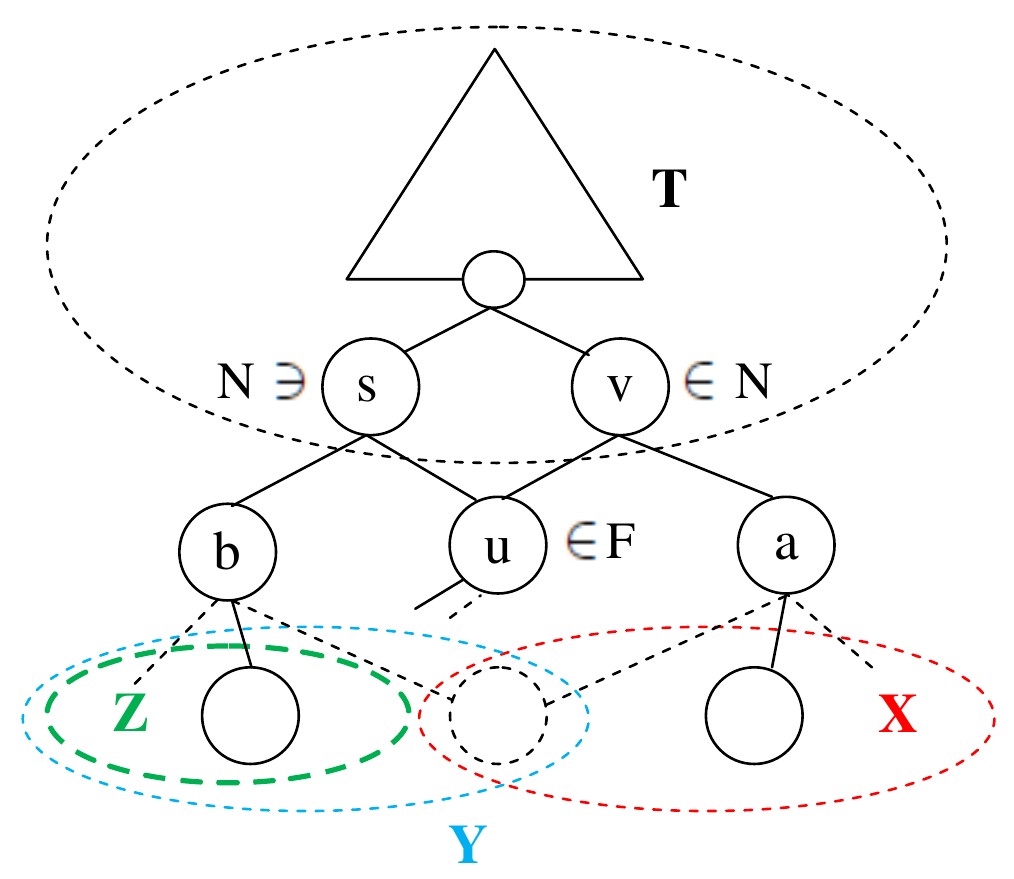}}
\caption{The case handled by Rules \ref{rule:commonF}--\ref{rule:36*}.}
\end{figure}

\begin{branchrule}{\normalfont 
[The same condition as in Rule \ref{rule:commonF}, except that there is $w\!\in\! Z$ s.t.~$\paths((N\setminus\{v,s\})\cup (X\setminus F),w,V\setminus(V_T\cup F\cup\{v,s,b\}))=\emptyset$.]
\begin{enumerate}
\item If \alg{Alg}$(T'=(V_T\cup \{a,u\}, E_T\cup\{(v,a),(v,u)\}),L\cup\{s\},M,F)$ accepts: Accept.
\item Else if \alg{Alg}$(T'=(V_T\cup \{u,b\}, E_T\cup\{(s,u),(s,b)\}),L\cup\{v\},M,F)$ accepts: Accept.
\item Return \alg{Alg}$(T'=(V_T\cup\{a,u,b\}\cup X\cup Y, E_T\cup\{(v,a),(v,u),(s,b)\}\cup\{(a,w): w\in X\}\cup\{(b,w): w\in Z\}),L,(M\cup X\cup Y)\setminus F,F)$.
\end{enumerate}
}\end{branchrule}

{\noindent For correctness, we need to show that in this rule, unlike the previous one, we can skip examining the instance in the first branch. Indeed, since there is $w\in Y\setminus X$ such that  $\paths((N\setminus\{v,s\})\cup (X\setminus F),w,V\setminus(V_T\cup F\cup\{v,s,b\}))=\emptyset$, once we determine that $v$ and $s$ are leaves, inserting $\{a,b\}$ to $F$ (see the first branch of the previous rule), we necessarily get a no-instance (since we cannot connect $w$ to the constructed tree). Observe that, although now there is $w\in Y\setminus X$ such that $\paths((N\setminus\{v,s,b\})\cup (X\setminus F),w,V\setminus(V_T\cup F\cup\{v,s,b\}))=\emptyset$, the dependency claim is still preserved since in the fourth branch, we also insert $Y\setminus F$ to $M$ (rather than only $X\setminus F$). Since $|X\cup Y|\geq 4$, the branching vector is at least as good as $(2,2,2)$, whose root is smaller than $3.188^{0.5}$.}

\begin{branchrule}\label{rule:36*}{\normalfont 
[There are $v\!\in\! N$, $s\!\in\!\siblings(v)\!\cap\! N$, $u\!\in\! F$ and $a,b$ s.t.~$\{a,u\}\!=\!\N(v)\!\setminus\! V_T$, $\{u,b\}\!=\!\N(s)\!\setminus\! V_T$, and ($|X\!\cup\! Y| \!\leq\!  3$ or $|X|\!\leq\!1$ or $Y\!\subseteq\! X$), where $X\!=\!\N(a)\!\setminus\!(V_T\!\cup\!\{v,u,b\})$ and $Y\!=\!\N(b)\!\setminus\!(V_T\!\cup\!\{s,u,a\})$.]
\begin{enumerate}
\item If \alg{Alg}$(T,L\cup\{v,s\},M,F\cup \{a,b\})$ accepts: Accept.
\item Else if \alg{Alg}$(T'=(V_T\cup \{a,u\}, E_T\cup\{(v,a),(v,u)\}),L\cup\{s\},M,F)$ accepts: Accept.
\item Return \alg{Alg}$(T'=(V_T\cup \{u,b\}, E_T\cup\{(s,u),(s,b)\}),L\cup\{v\},M,F)$.
\end{enumerate}
}\end{branchrule}

{\noindent This rule is similar to Rule \ref{rule:commonF}, except that now we do not examine its fourth branch. Recall that we established in Rule \ref{rule:commonF} that if $v$ and $s$ are internal vertices, so are $a$ and $b$ (where the children of $b$ do not include neighbors of $a$)---this is not possible if $Y\subseteq X$. Since $|X|\leq 1$ or $|X\cup Y|\leq 3$, we must have that one of $a$ or $b$ is a leaf or a vertex with only one child. Then, we obtain a contradiction in the same manner as in Rule \ref{rule:commonF}---although now $a$, for example, might be vertex of one child rather than a leaf, the proof is similar (we possibly need to reattach a vertex in the subtree of $a$ rather than $a$). The dependency claim is clearly preserved (in particular, recall again that we have already established (after Rule \ref{branch:Mafter}) that there is no $w\in V\setminus V_T$ such that $\paths(N\setminus\{v,s\},w,V\setminus(V_T\cup F))=\emptyset$). The branching vector is $(4,2,2)$, whose root is smaller than $3.188^{0.5}$.}

\subsection{Remaining Siblings in $N$}

{\bf Rule 37  was given in Section \ref{sec:examples}.}

\medskip
{\noindent Finally, we are only left with instances where there are $v\in N$, $s\in\siblings(v)\cap N$, $\{a,b\}=\N(v)\setminus V_T$ and $\{c,d\}=\N(s)\setminus V_T$ ($a,b,c,d$ are distinct vertices), such that $a,c\in F$ and $b,d\notin F$. Also, recall that there is no $w\in V\setminus V_T$ such that $\paths(N\setminus\{v,s\},w,V\setminus(V_T\cup F))=\emptyset$. These instances are handled in the two following rules.}

\begin{figure}[!h]\centering
\frame{\includegraphics[scale=0.5]{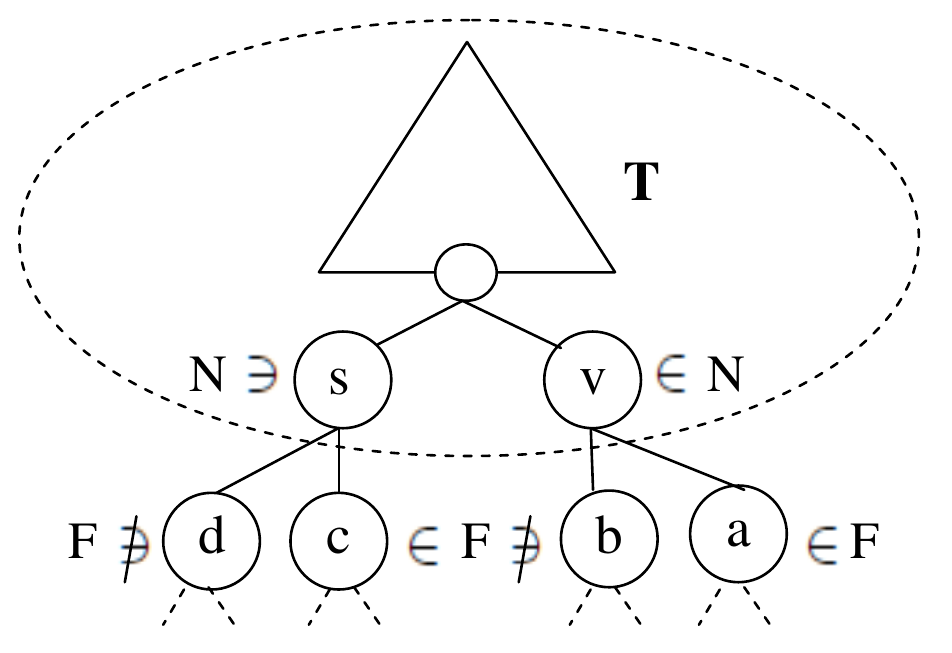}}
\caption{The case handled by Rules \ref{rule:38*} and \ref{rule:39*}.}
\end{figure}

\setcounter{reducerule}{37}

\begin{branchrule}\label{rule:38*}{\normalfont 
[There are $v,s,a,b,c,d$ as described in the remark above. Moreover, there is no $u\in V\setminus V_T$ such that $\paths(N\setminus\{v,s\},u,V\setminus(V_T\cup F\cup\{b\}))=\emptyset$ or $\paths(N\setminus\{v,s\},u,V\setminus(V_T\cup F\cup\{d\}))=\emptyset$.]
\begin{enumerate}
\item If \alg{Alg}$(T,L\cup\{v,s\},M,F\cup \{b,d\})$ accepts: Accept.
\item Else if \alg{Alg}$(T'=(V_T\cup\{a,b\}, E_T\cup\{(v,a),(v,b)\}),L\cup\{s,a\},M,F\setminus\{a\})$ accepts:~Accept.
\item Else if \alg{Alg}$(T'=(V_T\cup\{c,d\}, E_T\cup\{(s,c),(s,d)\}),L\cup\{v,c\},M,F\setminus\{c\})$ accepts:~Accept.
\item Return \alg{Alg}$(T'=(V_T\cup\{a,b,c,d\}, E_T\cup\{(v,a),(v,b),(s,c),(s,d)\}),L,M,F)$.
\end{enumerate}
}\end{branchrule}

{\noindent As in previous rules of the same form, this rule is exhaustive in the sense that we try all four options to determine the roles of $v$ and $s$, set the neighbors outside $T$ of an internal vertex as its children, and in the first branch (as in the first branch, e.g., of Rule \ref{branch:worse}), insert the neighbors outside $T$ of $v$ and $s$ to $F$. Also, as in Rule \ref{branch:worse}, the depednency claim is preserved in all branches. In the second branch, we have an instance where the only applicable next rule is one among Rules \ref{red:stop1}--\ref{red:stop3}, \ref{red:5}--\ref{rule:redF}, \ref{red:deg2deg2}, \ref{branch:neisinF} and \ref{branch:xsN3}--\ref{red:20} (in particular, Rule \ref{red:reach} is skipped because, by the condition of the rule, there is no $u\in V\setminus V_T$ such that $\paths(N\setminus\{v,s\},u,V\setminus(V_T\cup F\cup\{b\}))=\emptyset$). Thus, if the algorithm does not return a decision, we either apply a reduction rule where the measure decreases by at least $0.5$, or a branching rule whose branching vector is at least as good as $(1,2)$. The same claim applies for the instance in the third branch. Therefore, the branching vector is at least as good as $(4,2+(1,2),2+(1,2),2)=(4,4,4,3,3,2)$, whose root is smaller than $3.188^{0.5}$.}

\begin{branchrule}\label{rule:39*}{\normalfont 
[Remaining case. There are $a,b,c,d$ as described in the remark preceding the previous rule.]
\begin{enumerate}
\item If \alg{Alg}$(T'=(V_T\cup\{a,b\}, E_T\cup\{(v,a),(v,b)\}),L\cup\{s,a\},M,F\setminus\{a\})$ accepts:~Accept.
\item Else if \alg{Alg}$(T'=(V_T\cup\{c,d\}, E_T\cup\{(s,c),(s,d)\}),L\cup\{v,c\},M,F\setminus\{c\})$ accepts:~Accept.
\item Return \alg{Alg}$(T'=(V_T\cup\{a,b,c,d\}, E_T\cup\{(v,a),(v,b),(s,c),(s,d)\}),L,M,F)$.
\end{enumerate}
}\end{branchrule}

{\noindent This rule is similar to the previous one, except that we do not consider its first branch. However, there is no solution to the first branch (in the previous rule), since once we determine that $v,s,b,d$ are all leaves, we cannot extend the constructed tree to a spanning tree (since now there exists $u\in V\setminus V_T$ such that $\paths(N\setminus\{v,s,b\},u,V\setminus(V_T\cup F\cup\{b\}))=\emptyset$ or $\paths(N\setminus\{v,s,d\},u,V\setminus(V_T\cup F\cup\{d\}))=\emptyset$). As in the previous rule, the dependency claim is preserved. The branching vector is $(2,2,2)$, whose root is smaller than $3.188^{0.5}$.

\end{document}